\documentclass[epjST]{svjour}

\usepackage{bbold,graphics}

\usepackage{blindtext,hyperref}

\newcommand{\ab}{\bar a}
\newcommand{\ah}{\hat a}

\newcommand{\cB}{{\cal B}}

\newcommand{\cO}{{\cal O}}
\newcommand{\as}{\alpha_s}
\newcommand{\wh}{\widehat}
\newcommand{\nn}{\nonumber}

\newcommand{\um}{\mathbb{1}}
\newcommand{\wt}{\widetilde}

\newcommand{\IM}{\mbox{\rm Im}}

\newcommand{\eqn}[1]{(\ref{#1})}

\newcommand{\tvs}{\vbox{\vskip 6mm}}
\newcommand{\mvs}{\vbox{\vskip 8mm}}

\newcommand{\MSb}{{\overline{\rm MS}}}

\newcommand{\sfrac}[2]{\mbox{$\frac{#1}{#2}$}}

\begin{document}

\title{Higher-order behaviour of two-point current correlators}

\author{Matthias Jamin\thanks{\email{matthias.jamin@gmail.com}}}

\institute{University of Vienna, Faculty of Physics, Boltzmanngasse 5,
 A-1090 Wien, Austria}

\abstract{
Estimates of higher-order contributions for perturbative series in QCD, in
view of their asymptotic nature, are delicate, though indispensable for a
reliable error assessment in phenomenological applications. In this work,
the Adler function and the scalar correlator are investigated, and models
for Borel transforms of their perturbative series are constructed, which
respect general constraints from the operator product expansion and the
renormalisation group. As a novel ingredient, the QCD coupling is employed
in the so-called $C$-scheme, which has certain advantages. For the Adler
function, previous results obtained directly in the $\overline{\rm MS}$ scheme
are supported. Corresponding results for the scalar correlation function are
new. It turns out that the substantially larger perturbative corrections for
the scalar correlator in $\overline{\rm MS}$ are dominantly due to this scheme
choice, and can be largely reduced through more appropriate renormalisation
schemes, which are easy to realise in the $C$-scheme.
}

\maketitle

\section{Introduction}\label{intro}
In Quantum Field Theory, perturbative expansions in the coupling are generally
assumed to be divergent, being asymptotic at best. For QCD this is backed up
by studies of models and calculations in the large-$\beta_0$ approximation
\cite{gzj90,ben98}.\footnote{For historical reasons, we shall speak about the
``large-$\beta_0$'' approximation, although in the notation employed in this
work, the leading coefficient of the $\beta$-function is termed $\beta_1$.} In
phenomenological applications, two-point current correlation functions, like
the Adler function \cite{adl74} or the scalar correlator, play an important
role. Reliable estimates of uncertainties resulting from higher-order
corrections necessitate a good understanding of the behaviour of those
quantities at large perturbative order.

For an asymptotic expansion, within a radius of convergence, the Borel transform
in the coupling turns out to be convergent. For this reason, models for the
Borel transform which are based on physical constraints, like analyticity or
the renormalisation group (RG), can be important tools for gaining a better
insight into the large-order behaviour of basic QCD quantities, like the
current correlators. For the case of the QCD Adler function such a model
was first proposed in ref.~\cite{bj08}, and further investigated in more detail
in ref.~\cite{bbj12}.

In this contribution, the basic form of Borel models for the Adler function,
which are based on the renormalon structure, are reviewed. As a new ingredient,
the presentation shall be explicated in the $C$-scheme~\cite{bjm16,jm16},
which, as we shall see, has certain advantages. As a further novel addition, in
this article also Borel models for the scalar correlation function shall be
introduced and discussed.

To this end, in section~2, the QCD coupling and quark mass in the $C$-scheme
shall be defined. Section~3 then summarises our current knowledge of the
perturbative expansions of Adler function and scalar correlator. Next, in
section~4, the general form of renormalon pole terms for the Borel transform
of current correlators are derived from the RG and further ingredients, that
are required for the phenomenological investigation, are introduced. Central
results of the phenomenological applications are then presented in section~5,
and in section~6, some concluding remarks summarise this presentation.

\section{The \boldmath{$C$}-scheme}\label{sect2}

To keep this article self-contained, in the next two sub-sections, useful
relations for the QCD coupling as well as the quark mass in the $C$-scheme
are collected.

\begin{boldmath}
\subsection{QCD coupling in the $C$-scheme}\label{sect2.1}
\end{boldmath}
\noindent
The $C$-scheme coupling in QCD was introduced as well as applied to the
vector correlator in ref.~\cite{bjm16}, and will be employed in the following.
A second application to the scalar correlation function was worked out in
ref.~\cite{jm16}. The defining relation for the $C$-scheme coupling $\ah_Q^C$
takes the form
\begin{equation}
\label{ahat}
\frac{1}{\hat a_Q^C} + \frac{\beta_2}{\beta_1} \ln\hat a_Q^C -
\frac{\beta_1}{2}\,C \,\equiv\, \beta_1 \ln\frac{Q}{\Lambda^{\!\rm{RS}}}
\,=\, \frac{1}{a_Q^{\rm{RS}}} + \frac{\beta_2}{\beta_1}\ln a_Q^{\rm{RS}} -
\beta_1 \!\int\limits_0^{a_Q^{\rm{RS}}}\,
\frac{{\rm d}a}{\tilde\beta^{\rm{RS}}(a)} \,,
\end{equation}
where on the right-hand side $a_Q^{\rm{RS}}\equiv\alpha_s^{\rm{RS}}(Q)/\pi$ is
the QCD coupling in a particular renormalisation scheme RS, for example the
$\MSb$ scheme \cite{bbdm78}, and
\begin{equation}
\frac{1}{\tilde\beta^{\rm{RS}}(a)} \,\equiv\, \frac{1}{\beta^{\rm{RS}}(a)} -
\frac{1}{\beta_1 a^2} + \frac{\beta_2}{\beta_1^2 a} \,.
\end{equation}
In our conventions, the QCD $\beta$-function, together with its perturbative
expansion, is defined by
\begin{equation}
\label{aQRGE}
- Q\,\frac{{\rm d}a_Q^{\rm{RS}}}{{\rm d}Q} \,\equiv\,
\beta^{\rm{RS}}(a_Q^{\rm{RS}}) \,=\, \beta_1 \,(a_Q^{\rm{RS}})^2 +
\beta_2 \,(a_Q^{\rm{RS}})^3 + \beta_3^{\rm{RS}} \,(a_Q^{\rm{RS}})^4 + \ldots\,.
\end{equation}
Note that the first two $\beta$-coefficients $\beta_1$ and $\beta_2$ are
independent of the renormalisation scheme. Next, $\Lambda^{\!\rm{RS}}$ denotes
the $\Lambda$-parameter of QCD in the RS scheme. It coincides with the
$\Lambda$-parameter of the $\ah_Q^C$ coupling in the $C$-scheme with $C=0$,
which in the following shall be abbreviated by $\ab_Q\equiv \ah_Q^{C=0}$.
Even though the $C$-scheme coupling $\ah_Q^C$ also depends on the initial
renormalisation scheme of $a_Q^{\rm{RS}}$, for notational simplicity we refrain
from making this dependence explicit. The respective dependence can always be
compensated by a corresponding trivial shift in $C$.

An explicit solution of $\ah_Q^C$ as a function of the $\Lambda$-parameter can
be found in terms of the Lambert $W$-function \cite{wsdb18}. Defining
\begin{equation}
\label{LambdaC}
\Lambda^{\!C} \,\equiv\, \Lambda^{\!\rm{RS}} \,{\rm e}^{-C/2}
\quad {\rm and} \quad z \,\equiv\, -\,\frac{\beta_1}{\beta_2} \,\exp
\biggl( -\,\frac{\beta_1^2}{\beta_2} \ln\frac{Q}{\Lambda^{\!C}} \biggr)
\,=\, -\,\frac{\beta_1}{\beta_2} \biggl( \frac{\Lambda^{\!C}}{Q}
\biggr)^{\beta_1^2/\beta_2} \,,
\end{equation}
one obtains
\begin{equation}
\ah_Q^C \,=\, -\,\frac{\beta_1}{\beta_2\,W_{-1}(z)} \,,
\end{equation}
with $W(z)$ being a solution of the equation $W\!\exp(W)=z$ and the index
``$-1$'' signifies the proper branch of $W(z)$ to yield a physical coupling
with the correct properties. For further details the reader is referred to
ref.~\cite{wsdb18}.

From eq.~\eqn{ahat}, renormalisation group equations with respect to both
scale and scheme transformations can readily be derived. Simply taking the
derivatives with respect to $Q$ or $C$, one obtains
\begin{equation}
\label{CRGE}
- Q\,\frac{{\rm d}\ah_Q^C}{{\rm d}Q} \,\equiv\, \hat\beta\big(\ah_Q^C\big) \,=\,
\frac{\beta_1 (\ah_Q^C)^2}{\left(1 - \sfrac{\beta_2}{\beta_1}\, \ah_Q^C\right)}
\,=\, -\,2\,\frac{{\rm d}\ah_Q^C}{{\rm d}C} \,,
\end{equation}
where $\hat\beta\big(\ah_Q^C\big)$ is the simple, scheme-invariant
$\beta$-function\footnote{It only depends on the scheme-invariant
$\beta$-function coefficients $\beta_1$ and $\beta_2$.} corresponding to the
coupling $\ah_Q^C$, which can be given in closed form. From eq.~\eqn{CRGE},
one also observes that a change in the renormalisation scheme is equivalent to
an evolution in the renormalisation scale. A shift in the scale from $Q$ to
$\mu$ can be compensated by a transformation in the scheme from $C$ to $C'$,
satisfying the relation
\begin{equation}
\frac{Q^2}{\mu^2} \,=\, {\rm e}^{C-C'} \,.
\end{equation}
In the $C$-scheme both transformations are interchangeable and completely
equivalent.

This opens the possibility to arrive at the perturbative expansion in the
$C$-scheme at arbitrary $C$ by first computing the expansion in $\ah_Q^C$ at
$C=0$, and then employing the evolution equation to arrive at an arbitrary
$C$. Once again, this is completely analogous to the possibility of rederiving
the scale logarithms from the RGE in the renormalisation scale. Then, from
eq.~\eqn{ahat} the perturbative relation between the coupling $a_Q^{\rm{RS}}$
in an arbitrary scheme, for example the $\MSb$ scheme, and $\ab_Q$ is found
to be
\begin{eqnarray}
\label{afunab}
a_Q^{\rm{RS}} \,=\, \ab_Q \!&+&\! \biggl( \frac{\beta_3^{\rm RS}}{\beta_1} -
\frac{\beta_2^2}{\beta_1^2} \biggr) \ab_Q^3 + \biggl(
\frac{\beta_4^{\rm RS}}{2\beta_1} - \frac{\beta_2^3}{2\beta_1^3} \biggr) \ab_Q^4
\nn \\
\mvs
&+&\! \biggl( \frac{\beta_5^{\rm RS}}{3\beta_1} -
\frac{\beta_2\beta_4^{\rm RS}}{6\beta_1^2} +
\frac{5(\beta_3^{\rm RS})^2}{3\beta_1^2} -
\frac{3\beta_2^2\beta_3^{\rm RS}}{\beta_1^3} +
\frac{7\beta_2^4}{6\beta_1^4} \biggr) \ab_Q^5 + {\cal O}(\ab_Q^6) \,.
\end{eqnarray}
The higher $\beta$-function coefficients $\beta_n^{\rm RS}$ with $n\geq3$ on the
right-hand side are to be taken in the scheme corresponding to $a_Q^{\rm{RS}}$.
The relation between the coupling $\ah_Q^C$ at arbitrary $C$ and $\ab_Q$ can be
deduced by integrating the RGE \eqn{CRGE}, which results in the perturbative
expansion
\begin{eqnarray}
\label{abfunah}
\ab_Q \,=\, \ah_Q^C \!&+&\! \frac{\beta_1}{2}\,C\,(\ah_Q^C)^2 + \biggl(
\frac{\beta_2}{2}\,C + \frac{\beta_1^2}{4}\,C^2 \biggr)(\ah_Q^C)^3 \nn \\
\mvs
&+&\! \biggl( \frac{\beta_2^2}{2\beta_1}\,C + \frac{5\beta_1\beta_2}{8}\,C^2
+ \frac{\beta_1^3}{8}\,C^3 \biggr)(\ah_Q^C)^4 \nn \\
\mvs 
&+&\!\! \biggl( \frac{\beta_2^3}{2\beta_1^2}\,C + \frac{9\beta_2^2}{8}\,C^2 +
\frac{13\beta_1^2\beta_2}{24}\,C^3 + \frac{\beta_1^4}{16}\,C^4 \biggr)
(\ah_Q^C)^5 +{\cal O}\big((\ah_Q^C)^6\big) \,.
\end{eqnarray}

Eqs. \eqn{afunab} and \eqn{abfunah} complete the set of relations which
are needed in order to rewrite perturbative QCD expansions in terms of the
$C$-scheme coupling $\ah_Q^C$.

\begin{boldmath}
\subsection{Quark mass in the $C$-scheme}\label{sect2.2}
\end{boldmath}
\noindent
In minimal subtraction schemes, analogously to the scale dependence, also the
scheme dependence of the quark mass only originates from the QCD coupling.
Hence, the scheme evolution of a generic running $C$-scheme quark mass
$m_Q^C\equiv m^C(Q)$ is found to be
\begin{equation}
\frac{1}{m_Q^C} \frac{{\rm d}m_Q^C}{{\rm d}C} \,=\,
\frac{{\rm d}\ah_Q^C}{{\rm d}C}\,\frac{{\rm d}Q}{{\rm d}\ah_Q^C}\,
\frac{1}{m_Q^C} \frac{{\rm d}m_Q^C}{{\rm d}Q} \,=\,
-\,\frac{1}{2}\,\hat\gamma_m(\ah_Q^C) \,.
\end{equation}
The definition of the mass anomalous dimension, together with its first five
coefficients in the $\MSb$-scheme, is provided in appendix~\ref{appA}. Because
both scale and scheme dependences of the $C$-scheme coupling are given by the
$\beta$-function, this dependence cancels. Hence, similarly to the scheme
evolution of the coupling $\ah_Q^C$ which is given by the $\beta$-function,
scheme dependence of $m_Q^C$ is just governed by the quark-mass anomalous
dimension, expressed as a function of $\ah_Q^C$.  

This entails the additional finding that in the $C$-scheme an expression
involving the quark mass that is scale invariant, automatically also is scheme
invariant. Let us demonstrate that explicitly for the case of the invariant
quark mass $\wh m$, which is defined by
\begin{equation}
\label{mhat}
\wh m \,\equiv\, m_Q^C \,[\hat\alpha_s^C(Q)]^{-\gamma_m^{(1)}/\beta_1}
\exp\Biggl\{ \int\limits_0^{\ah_Q^C} \!{\rm d}\ah \biggl[
\frac{\gamma_m^{(1)}}{\beta_1\ah} - \frac{\hat\gamma_m(\ah)}{\hat\beta(\ah)}
\biggr] \Biggr\} \,.
\end{equation}
Taking the derivative of $\wh m$ with respect to $C$, employing product and
chain rules, yields
\begin{equation}
\frac{{\rm d}\wh m}{{\rm d}C} \,=\, \Biggl\{ -\,\frac{1}{2}\,
\hat\gamma_m(\ah_Q^C) + \frac{\gamma_m^{(1)}}{2\beta_1}\,\hat\beta(\ah_Q^C) -
\frac{1}{2}\,\hat\beta(\ah_Q^C) \biggl( \frac{\gamma_m^{(1)}}{\beta_1\ah_Q^C} -
\frac{\hat\gamma_m(\ah_Q^C)}{\hat\beta(\ah_Q^C)} \biggr) \Biggr\}\,\wh m
\,=\, 0 \,,
\end{equation}
which proves the claimed scheme independence of $\wh m$.

In order to relate the coefficients of the quark-mass anomalous dimension in
the $C$-scheme and another minimal renormalisation scheme, for example $\MSb$,
we still have to fix the global normalisation of the quark mass. This can be
done by assuming that the quark mass at a particular $C_m$ coincides with the
running quark mass $m_Q$ of a generic quark flavour in the $\MSb$-scheme, that
is
\begin{equation}
\label{mQCm}
m_Q^{\MSb} \,\equiv\, m_Q^{C_m} \,.
\end{equation}
This should always be possible since with a scheme evolution in $C$, any
value of the quark mass can be reached. Now equating the invariant mass in the
$\MSb$-scheme and in the $C$-scheme, as well as employing the normalisation of
eq.~\eqn{mQCm}, we can extract the coefficients of the quark-mass anomalous
dimension in the $C$-scheme. In particular, the first three coefficients are
found to be:
\begin{eqnarray}
\label{ghat}
\hat\gamma_m^{(1)} &=& \gamma_m^{(1)} \,, \quad
\hat\gamma_m^{(2)} \,=\, \gamma_m^{(2)} +
                         \frac{\beta_1}{2}\,\gamma_m^{(1)}\,C_m \,, \nn \\
\mvs
\hat\gamma_m^{(3)} &=& \gamma_m^{(3)} + \biggl( \frac{\beta_3}{\beta_1} -
\frac{\beta_2^2}{\beta_1^2} \biggr) \gamma_m^{(1)} + \frac{1}{2}\Big( \beta_2
\gamma_m^{(1)} + 2\beta_1\gamma_m^{(2)} \Big) C_m + \frac{\beta_1^2}{4}\,
\gamma_m^{(1)}\,C_m^2 \,.
\end{eqnarray}
The first relation is evident as the leading coefficient of the mass anomalous
dimension is scheme invariant. The renormalisation group coefficients without
hat are taken to be in the $\MSb$-scheme. Since the choice of $C_m$ is
arbitrary, we can simplify the relations by assuming equality with the $\MSb$
mass at $C_m=0$. Analogously to the coupling, denoting the corresponding
anomalous dimension with a bar, one obtains
\begin{equation}
\label{gbar}
\bar\gamma_m^{(1)} \,=\, \gamma_m^{(1)} \,, \quad
\bar\gamma_m^{(2)} \,=\, \gamma_m^{(2)} \,, \quad
\bar\gamma_m^{(3)} \,=\, \gamma_m^{(3)} + \biggl( \frac{\beta_3}{\beta_1} -
\frac{\beta_2^2}{\beta_1^2} \biggr) \gamma_m^{(1)} \,.
\end{equation}
Given the mass anomalous dimension together with the normalisation \eqn{mQCm},
the quark mass in the $C$-scheme is unambiguously defined.

\vspace*{0.5cm}
\section{QCD two-point correlation functions}\label{sect3}

In the following two sub-sections, our present knowledge of the perturbative
expansions of the Adler function and the scalar correlation function shall be
summarised.

\subsection{The vector two-point correlator}\label{sect3a}

The vector correlation function $\Pi_{\mu\nu}(p)$ in momentum space is
defined as
\begin{equation}
\label{Pimunu}
\Pi_{\mu\nu}(p) \,\equiv\,  i\!\int \! {\rm d}x \, e^{ipx} \,
\langle\Omega|\,T\{ j_\mu(x)\,j_\nu(0)^\dagger\}|\Omega\rangle\,,
\end{equation}
where $|\Omega\rangle$ denotes the physical QCD vacuum.
In order to avoid complications with so-called singlet diagrams, arising from
self-contractions of the quarks in the current, the normal-ordered current
$j_\mu(x)$ is taken to be flavour non-diagonal with the particular choice
\begin{equation}
\label{jvec}
j_\mu(x) \,=\, :\!\bar u(x)\gamma_\mu s(x)\!: \,,
\end{equation}
which for example plays a role in hadronic decays of the $\tau$ lepton into
strange final states. The correlator $\Pi_{\mu\nu}(p)$ admits the Lorentz
decomposition
\begin{eqnarray}
\label{PimunuDecomp}
\Pi_{\mu\nu}(p) &=& (p_\mu p_\nu - g_{\mu\nu}p^2)\,\Pi^{(1)}(p^2) +
p_\mu p_\nu\,\Pi^{(0)}(p^2) \nn \\
\tvs
&=& (p_\mu p_\nu - g_{\mu\nu}p^2)\,\Pi^{(1+0)}(p^2) +
g_{\mu\nu}\,p^2\,\Pi^{(0)}(p^2) \,,
\end{eqnarray}
where the superscripts denote the components corresponding to angular momentum
$J=1$ (transversal) and $J=0$ (longitudinal) in the hadronic rest frame. In the
second way of writing eq.~\eqn{PimunuDecomp}, the Lorentz-scalar correlators
$\Pi^{(1+0)}(p^2)$ and $p^2\,\Pi^{(0)}(p^2)$ are free from kinematical
singularities at $p^2=0$. These are the correlation functions that will be
investigated in the following.

The correlator $\Pi^{(0)}(s)$ with $s\equiv p^2$ turns out to be proportional
to the quark masses, and hence vanishes in the limit of massless quarks.
It will be further discussed in the next sub-section. For the correlator
$\Pi^{(1+0)}(s)$, the purely perturbative expansion assumes the general
structure
\begin{equation}
\label{Pi10exp}
\Pi^{(1+0)}_{\rm PT}(s) \,=\, -\,\frac{N_c}{12\pi^2} \sum\limits_{n=0}^\infty
a_\mu^n \sum\limits_{k=0}^{n+1} c_{n,k}\,L^k  \,, \quad L\,\equiv\,
\ln\frac{-s}{\mu^2} \,,
\end{equation}
with $a_\mu\equiv a(\mu^2) \equiv\as(\mu)/\pi$ and $\mu$ the renormalisation
scale. $\Pi^{(1+0)}(s)$ itself is not a physical quantity in the sense that
it contains a renormalisation scale and scheme dependent subtraction constant.
This subtraction constant can either be removed by taking the imaginary part
which corresponds to the spectral function
$\rho^{(1+0)}(s)\equiv \IM\,\Pi^{(1+0)}(s+i 0)/\pi$, or by taking a derivative
with respect to $s$, which leads to the Adler function
\begin{equation}
\label{Ds}
D(s) \,\equiv\, -\,s\,\frac{\rm d}{{\rm d}s}\,\Pi^{(1+0)}(s) \,.
\end{equation}
Both, the spectral function and the Adler function are physical in the above
mentioned sense. However, let us remark that the natural domain of the spectral
function is for real and Minkowskian $s>0$, while that of the Adler function
is for Euclidean $s<0$. Nonetheless, apart from the cut on the positive real
axis, the Adler function can be continued into the whole complex $s$-plane,
and the two functions are related by
\begin{equation}
\label{specfunD}
\rho^{(1+0)}(s) \;=\; \frac{1}{2\pi i}\!\int\limits_{s+i0}^{s-i0}
\frac{D(s')}{s'}\, {\rm d}s' \,.
\end{equation}

From eq.~\eqn{Pi10exp}, the general perturbative expansion of the Adler
function follows as
\begin{equation}
\label{Dexp}
D_{\rm PT}(s) \,=\, \frac{N_c}{12\pi^2} \sum\limits_{n=0}^\infty a_\mu^n
\sum\limits_{k=1}^{n+1} k\, c_{n,k}\,L^{k-1} \,.
\end{equation}
In this expression, only the coefficients $c_{n,1}$ have to be considered as
independent. The coefficients $c_{n,k}$ with $k=2,\ldots,n$ can be related to
the $c_{n,1}$ and $\beta$-function coefficients by means of the renormalisation
group equation (RGE), while the coefficients $c_{n,0}$ do not appear in
measurable quantities and $c_{n,n+1}=0$ for $n\geq 1$. Up to order $\as^4$,
the respective RG constraints are provided in eq.~(2.11) of ref.~\cite{bj08}.

Since the Adler function $D(s)$ satisfies a homogeneous RGE, the logarithms in
eq.~\eqn{Dexp} can be summed with the choice $\mu^2=-s\equiv Q^2$, leading
to the simple expression
\begin{equation}
\label{Dresum}
D_{\rm PT}(Q^2) \,=\, \frac{N_c}{12\pi^2} \sum\limits_{n=0}^\infty
c_{n,1}\,a_Q^n \,,
\end{equation}
where $a_Q\equiv\as(Q)/\pi$. The independent coefficients $c_{n,1}$ are known
analytically up to order $\as^4$ \cite{gkl91,ss91,bck08}. At $N_c=3$ and
$N_f=3$, in the $\MSb$-scheme, they read:
\begin{eqnarray}
\label{cn1}
c_{0,1} &=& c_{1,1} \,=\, 1 \,, \quad
c_{2,1} \,=\, \sfrac{299}{24} - 9\zeta_3 \,=\, 1.640 \,, \nn \\
\mvs
c_{3,1} &=& \sfrac{58057}{288} - \sfrac{779}{4}\zeta_3 + \sfrac{75}{2} \zeta_5
\,=\, 6.371 \,, \\
\mvs
c_{4,1} &=& \sfrac{78631453}{20736} - \sfrac{1704247}{432}\zeta_3 +
\sfrac{4185}{8}\zeta_3^2 + \sfrac{34165}{96}\zeta_5 - \sfrac{1995}{16}\zeta_7
\,=\, 49.076 \,. \nn
\end{eqnarray}

As a final preparation for the subsequent analysis, employing eq.~\eqn{afunab},
we rewrite the expansion of the Adler function in terms of the $C$-scheme
coupling $\ab_Q$ at $C=0$. It will furthermore be convenient to introduce the
reduced Adler function $\hat D_{\rm PT}(a_Q)$ in the following way:
\begin{eqnarray}
\label{Dhat}
\hat D_{\rm PT}(a_Q) &\equiv& 4\pi^2 D_{\rm PT}(a_Q) - 1 \,=\,
\sum_{n=1}^\infty c_{n,1} a_Q^n \,\equiv\, \sum_{n=1}^\infty
\bar c_{n,1} \ab_Q^n \nn \\
\mvs
&=& \ab_Q + 1.640\,\ab_Q^2 + 7.682\,\ab_Q^3 + 61.06\,\ab_Q^4 +\ldots \,,
\end{eqnarray}
which also defines the independent perturbative coefficients $\bar c_{n,1}$.
Only the coefficients $\bar c_{3,1}$ and $\bar c_{4,1}$ turn out different from
the $\MSb$ coefficients, and analytically as well as numerically read:
\begin{eqnarray}
\label{cb3cb4}
\bar c_{3,1} &=& \sfrac{262955}{1296} - \sfrac{779}{4}\zeta_3 +
                 \sfrac{75}{2} \zeta_5 \,=\, 7.682 \,, \nn \\
\tvs
\bar c_{4,1} &=& \sfrac{357259199}{93312} - \sfrac{1713103}{432}\zeta_3 +
                 \sfrac{4185}{8}\zeta_3^2 + \sfrac{34165}{96}\zeta_5 -
                 \sfrac{1995}{16}\zeta_7 \,=\, 61.060 \,.
\end{eqnarray}
The corresponding expansion in terms of the $C$-scheme coupling $\ah_Q$ for
arbitrary $C$ can be obtained by inserting the relation \eqn{abfunah} into
eq.~\eqn{Dhat}. The resulting expression then coincides with eq.~(12) of
ref.~\cite{bjm16} or eq.~(15) of ref.~\cite{bjm17}.

\subsection{The scalar two-point correlator}\label{sect3b}

As the second correlation function, the scalar two-point correlator $\Psi(p^2)$
is introduced, which is defined by
\begin{equation}
\label{Psi}
\Psi(p^2) \,\equiv\, i\!\int\!{\rm d}x \,{\rm e}^{ipx} \langle\Omega|
T\{j(x) j^\dagger(0)\}|\Omega\rangle \,.
\end{equation}
For our application, the scalar current $j(x)$ is chosen to arise from the
divergence of the vector current,
\begin{equation}
\label{jtau}
j(x) \,=\, \partial^\mu \!:\!\bar u(x)\gamma_\mu s(x)\!: \;=\,
i\,(m_u-m_s) \!:\!\bar u(x) s(x)\!: \,.
\end{equation}
This choice has the advantage of an additional factor of the quark masses,
which makes the currents $j(x)$ renormalisation group invariant (RGI).

The purely perturbative expansion of $\Psi(p^2)$ is known up to order $\as^4$
and takes the general form
\begin{equation}
\label{PsiPT}
\Psi_{\rm PT}(s) \,=\, -\,\frac{N_c}{8\pi^2} \,m_\mu^2 \,s
\sum\limits_{n=0}^\infty a_\mu^n \sum\limits_{k=0}^{n+1} d_{n,k} L^k \,.
\end{equation}
To simplify the notation, we have introduced the generic mass factor $m_\mu$
which stands for the combination $(m_u(\mu)-m_s(\mu))$. The running quark
masses and the QCD coupling are renormalised at the common scale $\mu$, which
enters in the logarithm $L=\ln(-s/\mu^2)$. As a matter of principle, different
scales could be introduced for the renormalisation of coupling and quark
masses. We shall return to a discussion of this aspect below.

At each perturbative order $n$, the only independent coefficients $d_{n,k}$
are the $d_{n,1}$. The coefficients $d_{n,0}$ depend on the renormalisation
prescription and do not contribute in physical quantities, while all remaining
coefficients $d_{n,k}$ with $k>1$ can again be obtained by means of the
renormalisation group equation (RGE). To order $\as^4$ they are listed in
eq.~(A.6) of ref.~\cite{jm16}. The normalisation in eq.~\eqn{PsiPT} is chosen
such that $d_{0,1}=1$. Again, setting $N_c=3$ and $N_f=3$, as well as employing
the $\MSb$-scheme, the coefficients $d_{n,1}$ up to $\cO(\as^4)$ were found to
be \cite{gkls90,che96,bck05}:
\begin{eqnarray}
\label{d01tod41}
d_{0,1} &=& 1 \,, \qquad
d_{1,1} \,=\, \sfrac{17}{3} \,=\, 5.6667 \,, \qquad
d_{2,1} \,=\, \sfrac{9631}{144} - \sfrac{35}{2} \zeta_3 \,=\, 45.846 \,,
\nn \\
\mvs
d_{3,1} &=& \sfrac{4748953}{5184} - \sfrac{91519}{216} \zeta_3 -
\sfrac{5}{2} \zeta_4 + \sfrac{715}{12} \zeta_5 \,=\, 465.85 \,, \\
\mvs
d_{4,1} &=& \sfrac{7055935615}{497664} - \sfrac{46217501}{5184} \zeta_3 +
\sfrac{192155}{216} \zeta_3^2 - \sfrac{17455}{576} \zeta_4 +
\sfrac{455725}{432} \zeta_5 - \sfrac{625}{48} \zeta_6 -
\sfrac{52255}{256} \zeta_7 \,=\, 5588.7 \,. \nn
\end{eqnarray}

Like the vector correlator, the correlator $\Psi(s)$ itself is not related to a
measurable quantity. It grows linearly with $s$ as $s$ tends to infinity, and
hence has two unphysical subtraction constants which can be removed by taking
two derivatives with respect to $s$, such that $\Psi^{''}(s)$ is independent of
the renormalisation prescription. Let us also remark that $\Psi(s)$ is related
to $\Pi^{(0)}(s)$ through a Ward identity \cite{bro81}, leading to
\begin{equation}
\label{Pi0Psi}
s\,\Pi^{(0)}(s) \,=\, \frac{1}{s}\,\big[\, \Psi(s) - \Psi(0) \,\big] \,.
\end{equation}

Employing \eqn{PsiPT}, the general perturbative expansion of $\Psi^{''}(s)$
reads
\begin{equation}
\label{PsippPT}
\Psi_{\rm PT}^{''}(s) \,=\, -\,\frac{N_c}{8\pi^2}\,\frac{m_\mu^2}{s}\,
\sum\limits_{n=0}^\infty a_\mu^n \sum\limits_{k=1}^{n+1} d_{n,k} \,k
\,\big[ L^{k-1} + (k-1) L^{k-2} \big] \,.
\end{equation}
Like for the Adler function, being a physical quantity, $\Psi^{''}(s)$
satisfies a homogeneous RGE, and therefore the logarithms can be summed with
the scale choice $\mu^2=-s\equiv Q^2$, resulting in the compact expression
\begin{equation}
\label{Psippres}
\Psi_{\rm PT}^{''}(Q^2) \,=\, \frac{N_c}{8\pi^2}\,\frac{m_Q^2}{Q^2}\,\biggl\{
\, 1 + \sum\limits_{n=1}^\infty \,(d_{n,1} + 2 d_{n,2}) \,a_Q^n \,\biggr\} \,.
\end{equation}
In this way, both the running quark mass as well as the running QCD coupling
are to be evaluated at the renormalisation scale $Q$. The dependent
coefficients $d_{n,2}$ can be calculated from the RGE. A closed expression
is provided in eq.~\eqn{dn2} of appendix~\ref{appA}, together with the
coefficients of the QCD $\beta$-function and mass anomalous dimension.
Numerically, at $N_f=3$, the perturbative coefficients
$d_{n,1}^{\,''}\equiv d_{n,1} + 2 d_{n,2}$ of eq.~\eqn{Psippres} take the
$\MSb$ values
\begin{equation}
\label{dt11todt41n}
d_{1,1}^{\,''} \,=\,  3.6667 \,, \qquad
d_{2,1}^{\,''} \,=\, 14.179  \,, \qquad
d_{3,1}^{\,''} \,=\, 77.368  \,, \qquad
d_{4,1}^{\,''} \,=\, 511.83  \,.
\end{equation}
It is observed that the coefficients \eqn{dt11todt41n} for the physical
correlator are substantially smaller than the $d_{n,1}$ of eq.~\eqn{d01tod41},
but still much larger than the $c_{n,1}$ \eqn{cn1} for the Adler function.

For the ensuing investigation it will be advantageous to remove the running
effects of the quark mass from the remaining perturbative series. This can
be achieved by rewriting the running quark masses $m_q(\mu)$ in terms of RGI
quark masses $\wh m_q$ of eq.~\eqn{mhat}, this time expressed in terms of
$\MSb$ masses:
\begin{equation}
\label{mmumhat}
m_q(\mu) \,\equiv\, \wh m_q \,[\alpha_s(\mu)]^{\gamma_m^{(1)}/\beta_1}
\exp\Biggl\{ \int\limits_0^{a_\mu} \!{\rm d}a \biggl[
\frac{\gamma_m(a)}{\beta(a)} - \frac{\gamma_m^{(1)}}{\beta_1 a} \biggr]
\Biggr\} \,.
\end{equation}
Accordingly, we define a modified perturbative expansion with new coefficients
$r_n$,
\begin{equation}
\label{Psippmhat}
\Psi_{\rm PT}^{''}(Q^2) \,=\, \frac{N_c}{8\pi^2}\,
\frac{\wh m^2}{Q^2} \,[\alpha_s(Q)]^{2\gamma_m^{(1)}/\beta_1}
\biggl\{\, 1 + \sum_{n=1}^\infty \, r_n \,a_Q^n \,\biggr\} \,,
\end{equation}
which now contain contributions from the exponential factor in eq.~\eqn{mhat}.
At $N_f=3$ and in the $\MSb$-scheme, the coefficients $r_n$ take the values
\begin{eqnarray}
\label{r1tor4}
r_1 &=& \sfrac{442}{81} \,=\, 5.4568 \,, \qquad
r_2 \,=\, \sfrac{2449021}{52488} - \sfrac{335}{18} \zeta_3 \,=\, 24.287 \,,
\nn \\
\mvs
r_3 &=& \sfrac{24657869923}{51018336} - \sfrac{678901}{1944} \zeta_3 +
\sfrac{18305}{324} \zeta_5 \,=\, 122.10 \,, \\
\mvs
r_4 &=& \sfrac{378986482023877}{66119763456} - \sfrac{21306070549}{3779136}
\zeta_3 + \sfrac{601705}{648} \zeta_3^2 + \sfrac{445}{96} \zeta_4 +
\sfrac{3836150}{6561} \zeta_5 - \sfrac{3285415}{20736} \zeta_7 \,=\, 748.09 \,.
\nn
\end{eqnarray}
The order $\as^4$ coefficient $r_4$ depends on quark-mass anomalous dimensions
as well as $\beta$-function coefficients up to five-loops which for the
convenience of the reader in our conventions have been collected in
appendix~\ref{appA}. Let us remark that the $\zeta_4$ term that is present in
$d_{3,1}$ as well as $d_{3,1}^{\,''}$, and the $\zeta_6$ term being present in
$d_{4,1}$ as well as $d_{4,1}^{\,''}$, have been cancelled by the additional
contribution. The respective cancellation has also been observed in
ref.~\cite{bck17} for a related quantity.

As the last step, similarly to the preceding sub-section, we reexpress the
QCD coupling in terms of $\bar\alpha_s$. The perturbative expansion of
$\Psi_{\rm PT}^{''}$ then assumes the form
\begin{equation}
\label{Psippabar}
\Psi_{\rm PT}^{''}(Q^2) \,=\, \frac{N_c}{8\pi^2}\,
\frac{\wh m^2}{Q^2} \,[\bar\alpha_s(Q)]^{2\gamma_m^{(1)}/\beta_1}
\biggl\{\, 1 + \sum_{n=1}^\infty \, \bar r_n \,\ab_Q^n \,\biggr\} \,,
\end{equation}
defining the coefficients $\bar r_n$, which take the particular values
\begin{eqnarray}
\label{rb1torb4}
\bar r_1 &=& \sfrac{442}{81} \,=\, 5.4568 \,, \qquad
\bar r_2 \,=\, \sfrac{2510167}{52488} - \sfrac{335}{18} \zeta_3 \,=\, 25.452 \,,
\nn \\
\mvs
\bar r_3 &=& \sfrac{12763567259}{25509168} - \sfrac{673561}{1944} \zeta_3 +
\sfrac{18305}{324} \zeta_5 \,=\, 142.44 \,, \\
\mvs
\bar r_4 &=& \sfrac{49275071521973}{8264970432} - \sfrac{10679302931}{1889568}
\zeta_3 + \sfrac{601705}{648} \zeta_3^2 + \sfrac{117947335}{209952} \zeta_5 -
\sfrac{3285415}{20736} \zeta_7 \,=\, 932.71 \,. \nn
\end{eqnarray}
It is amusing to observe that now even the $\zeta_4$ term remaining in $r_4$
got cancelled by a corresponding contribution in $\beta_5$, originating from
the global $\bar\alpha_s$ prefactor, such that only odd-integer $\zeta$-function
contributions persist. Even though we have only provided results for $N_f=3$,
we have convinced ourselves that the cancellation of even $\zeta$ values is
in fact independent of the number of flavours. This observation is discussed
in substantially more detail and generalised to other quantities in
refs.~\cite{jm17,dv17,bc18}. In the following sections, we shall furthermore
investigate rewriting the prefactor and the remaining perturbative expansion
into a general $C$-scheme coupling $\ah_Q^C$, which provides an interesting
handle on reshuffling the series.

\section{Borel transforms in the \boldmath{$\ah$} coupling}\label{sect4}

\subsection{General renormalon-pole structure}
\label{sect4.1}

This chapter contains an extension of the discussion performed in section~5 of
ref.~\cite{bj08}, but closely follows the material already presented there and
in part in the review~\cite{ben98}. Consider the OPE of a physical quantity
$D(Q)$ which is assumed to be defined such that it is dimensionless. Physical
means that it does not depend on renormalisation scale nor renormalisation
scheme. Particular examples would be the Adler function of eq.~\eqn{Ds} or the
physical scalar correlator $\Psi^{''}(s)$ of eq.~\eqn{PsippPT}. The general
structure of the OPE for $D(Q)$, expressed in the coupling $\ah_Q$,\footnote{For
notational simplicity, in the ensuing discussion we have dropped the superscript
$C$ in the $C$-scheme coupling $\ah_Q^C$.} can suggestively be written as
\begin{equation}
\label{ope}
D(Q) \,=\, C_\um(\ah_Q) + \sum_{O_d} \wh C_{O_d}(\ah_Q)
\,\frac{\langle \wh O_d\rangle}{Q^d} \,.
\end{equation}
$C_\um$ corresponds to the purely perturbative term, or unit operator, and
the higher-dimensional contributions are expressed in terms of renormalisation
group invariant (RGI) operators $\wh O_d$. If at a given dimension a set of
operators contributes, the basis should be chosen such that the leading-order
anomalous dimension matrix is diagonal. The structure of dimension-6 operator
contributions to vector and axialvector correlators is, for example, examined
in ref.~\cite{bhj15}.

Expressing the generic structure of the different contributions to $D(Q)$
requires some explanation. In general, a given correlation function may admit
global powers $\delta$ of $a_Q$. As was already encountered above, for example
$\delta$ is non-vanishing for the scalar correlation function with
$\delta=2\gamma_m^{(1)}/\beta_1$, see eq.~\eqn{Psippabar}, or the scalar
gluonium correlation function \cite{jam12} (not discussed in this work) for
which $\delta=2$. Hence, we choose to factor out this global prefactor and
write the perturbative part $C_\um(\ah_Q)$ as
\begin{equation}
\label{C0aQ}
C_\um(\ah_Q) \,=\, C_\um^{(0)}\,[\ah_Q]^\delta \Big[\, 1+\wt C_\um(\ah_Q)
\,\Big] \,.
\end{equation}

Regarding the OPE, a generic higher-dimensional term takes the form
\begin{equation}
\label{Dimd}
\wh C_{O_d}(\ah_Q) \,\frac{\langle \wh O_d\rangle}{Q^d} \,=\,
C_{O_d}^{(0)} \,[\ah_Q]^\delta \,[\ah_Q]^\eta \Big[\, 1 +
\wt C_{O_d}^{(1)}\,\ah_Q + \wt C_{O_d}^{(2)}\,\ah_Q^2 + \ldots \,\Big]
\frac{\langle \wh O_d\rangle}{Q^d} \,,
\end{equation}
where again we have explicitly factored out $[\ah_Q]^\delta$, such that this
factor can be considered global to the full correlation function. The remaining
exponent $\eta$ in general receives contributions from explicit powers
$[\ah_Q]^n$ which may be present at leading order, and the leading-order
anomalous dimensions $\gamma_{O_d}^{(1)}$ of the operators. Lastly, the
scale-invariant operator $\wh O_d$ in eq.~\eqn{Dimd} is defined by
\begin{equation}
\label{Ohat}
\wh O_d \,\equiv\, O_d(\mu)\,\exp\biggl\{-\!\int \frac{\gamma_{O_d}(\ah_\mu)}
{\beta(\ah_\mu)}\,{\rm d}\ah_\mu \biggr\} \,,
\end{equation}
where the anomalous dimension $\gamma_{O_d}$ of the operator $O_d$ is given by
\begin{equation}
\label{Omu}
-\,\mu\,\frac{{\rm d}}{{\rm d}\mu}\,O_d(\mu) \,\equiv\, \gamma_{O_d}(\ah_\mu)\,
O_d(\mu) \,=\, \left[\, \gamma_{O_d}^{(1)}\,\ah_\mu +
\gamma_{O_d}^{(2)}\,\ah_\mu^2 + \gamma_{O_d}^{(3)}\,\ah_\mu^3 + \ldots
\,\right] O_d(\mu) \,.
\end{equation}
Since the operators~\eqn{Ohat} will only be needed up to a multiplicative
factor, we do not have to specify the constant of integration, and without
loss of generality, it can be assumed to be zero.

Next, we rewrite the $Q$-dependence of a higher-dimensional OPE contribution
in terms of $\ah_Q$. From eq.~\eqn{ahat}, we find
\begin{equation}
\label{LaoQd}
\biggl(\frac{\Lambda^{\!\rm{RS}}}{Q}\!\biggr)^{\!d} \,=\,
{\rm e}^{-\frac{d}{\beta_1 \ah_Q}+\frac{d}{2}C}\,
[\ah_Q]^{-d\frac{\beta_2}{\beta_1^2}} \quad \textrm{or} \quad
\biggl(\frac{\Lambda^{\!C}}{Q}\!\biggr)^{\!d} \,=\,
{\rm e}^{-\frac{d}{\beta_1 \ah_Q}}\,
[\ah_Q]^{-d\frac{\beta_2}{\beta_1^2}} \,,
\end{equation}
where in the second relation the $C$-dependence has been absorbed into the
$\Lambda$-parameter according to the definition in eq.~\eqn{LambdaC}. Employing
the second equation, the $Q$-dependent part of the operator contribution reads
\begin{eqnarray}
\frac{\wh C_{O_d}(\ah_Q)}{Q^d} &=&
\frac{\wh C_{O_d}(\ah_Q)}{(\Lambda^{\!C})^d}\,
{\rm e}^{-\frac{d}{\beta_1 \ah_Q}}
\,[\ah_Q]^{-d\frac{\beta_2}{\beta_1^2}} \nonumber \\
\mvs
\label{CdoQd}
&=& \frac{C_{O_d}^{(0)}}{(\Lambda^{\!C})^d}\,
{\rm e}^{-\frac{d}{\beta_1 \ah_Q}} \,[\ah_Q]^\delta \,
[\ah_Q]^{\eta-d\frac{\beta_2}{\beta_1^2}}
\Big[\, 1 + \wt C_{O_d}^{(1)}\,\ah_Q + \wt C_{O_d}^{(2)}\,\ah_Q^2 +
\ldots \,\Big] \,.
\end{eqnarray}

The aim is to compare the latter structure to potential exponentially
suppressed terms in the perturbative part. To proceed we express
$\wt C_\um(\ah_Q)$, defined in eq.~\eqn{C0aQ}, by means of a Borel
transformation,
\begin{equation}
\label{BC0t}
\wt C_\um(\ah_Q) \,\equiv\,
\int\limits_0^\infty\! {\rm d}t\, {\rm e}^{-t/\ah_Q} B[\wt C_\um](t) \,=\,
\frac{2}{\beta_1}\! \int\limits_0^\infty\! {\rm d}u\,
{\rm e}^{-\frac{2u}{\beta_1 \ah_Q}} B[\wt C_\um](u) \,,
\end{equation}
with $t=2u/\beta_1$. If $\wt C_\um(\ah_Q)$ admits the perturbative expansion
\begin{equation}
\label{C1tilde}
\wt C_\um(\ah_Q) \,=\, \sum\limits_{n=1}^\infty \wt C_\um^{(n)}\,\ah_Q^n \,,
\end{equation}
the expansion of the Borel transform $B[\wt C_\um](t)$ is given by
\begin{equation}
B[\wt C_\um](t) \,=\, \sum\limits_{n=0}^\infty \wt C_\um^{(n+1)}\,
\frac{t^n}{n!}\,.
\end{equation}
To find the Borel transform that matches the $Q$-dependence of~\eqn{CdoQd},
we take the following ansatz for a particular IR renormalon pole located at
$u=p$:
\begin{equation}
\label{BR3PIR}
B[\wt C_{\um,p}^{\rm IR}](u) \,\equiv\, \frac{d_p^{\rm IR}}{(p-u)^\gamma}\,
\Big[\, 1 + b_1 (p-u) + b_2 (p-u)^2 +\ldots \,\Big] \,.
\end{equation}
Defining the Borel integral \eqn{BC0t} by the principal-value prescription, and
employing eq.~(A.8) of ref.~\cite{bj08}, the imaginary ambiguity corresponding
to the Borel integral of $B[\wt C_{\um,p}^{\rm IR}](u)$ is found to be:
\begin{eqnarray}
\label{ImR3P}
\IM\Big[\wt C_{\um,p}^{\rm IR}(\ah_Q) \Big] &=&
\pm\, \biggl(\frac{2}{\beta_1}\biggr)^{\!\gamma} d_p^{\rm IR}
\sin(\pi\gamma)\,\Gamma(1-\gamma)\,{\rm e}^{-\frac{2p}{\beta_1 \ah_Q}}\,
(\ah_Q)^{1-\gamma} \nn \\
\mvs
&\times& \!\biggl[\, 1 + b_1\frac{\beta_1}{2} \,(\gamma-1) \,\ah_Q +
b_2\frac{\beta_1^2}{4} \,(\gamma-1)(\gamma-2)\,\ah_Q^2 + \ldots \,\biggr] .
\end{eqnarray}
Assuming that this ambiguity gets cancelled by a corresponding ambiguity
in the definition of the operator matrix elements, as well as comparing
eqs.~\eqn{CdoQd} and \eqn{ImR3P}, one readily deduces:
\begin{equation}
\label{gammat}
p \,=\, \frac{d}{2} \,, \quad
\gamma \,=\, 1 - \eta + 2p\,\frac{\beta_2}{\beta_1^2} \,, \quad
b_1 \,=\, \frac{2 \wt C_{O_d}^{(1)}}{\beta_1(\gamma-1)} \,, \quad
b_2 \,=\, \frac{4 \wt C_{O_d}^{(2)}}{\beta_1^2 \,(\gamma-1)(\gamma-2)} \,.
\end{equation}
Taylor expanding the ansatz~\eqn{BR3PIR} in $u$ and performing the Borel
integral term by term yields the perturbative series:
\begin{eqnarray}
\label{Ral3PIR}
\wt C_{\um,p}^{\rm IR}(\ah_Q) &=& \frac{d_p^{\rm IR}}
{p^\gamma \,\Gamma(\gamma)}\, \sum\limits_{n=0}^\infty\,
\Gamma(n+\gamma) \biggl(\frac{\beta_1}{2p}\biggr)^{\!n}
(\ah_Q)^{n+1} \nn \\
\mvs
&\times& \!\biggl[\, 1 + \frac{2 p}{\beta_1}
\frac{\wt C_{O_d}^{(1)}}{(n+\gamma-1)} + \biggl(\frac{2 p}{\beta_1}\biggr)^{\!2}
\frac{\wt C_{O_d}^{(2)}}{(n+\gamma-1)(n+\gamma-2)} +
{\cal O}\left(\frac{1}{n^3}\right) \,\biggr] .
\end{eqnarray}
Eq.~\eqn{Ral3PIR} expressed in the coupling $\ah_Q$ extends the corresponding
$\MSb$ coupling eq.~(3.51) of ref.~\cite{ben98} to include terms of order
$1/n^2$ in the large-order behaviour of the perturbative series.

\subsection{Modified Borel transform}
\label{sect4.2}

Employing a conventional definition of the Borel transformation like in
eq.~\eqn{BC0t} in full QCD entails that the Borel transform $B[\wt C_\um](t)$
has a non-trivial dependence on the renormalisation scheme for the coupling.
In the past, this motivated the introduction of a so-called ``modified'' Borel
transform \cite{byz92,gru93}, which is based on the QCD $\Lambda$-parameter,
and which shares with it the simple transformation properties under scheme
changes.

We concentrate here on a definition of the modified Borel transform which
is inspired by the work of Brown, Yaffe and Zhai \cite{byz92} (see also
ref.~\cite{bo20}), and takes the form 
\begin{equation}
\label{mBTMJ}
\cB[\wt C_\um](t) \,\equiv\, \sum\limits_{n=1}^\infty \,
\frac{n\,\Gamma(1+\lambda\,t)}{\Gamma(n+1+\lambda\,t)}\,\wt C_\um^{(n)}
t^{n-1} \,,
\end{equation}
with $\lambda\equiv\beta_2/\beta_1$.
Similar and related modified Borel transforms were also considered in the
literature, but we shall not discuss them here in more detail, since the
practical application of the modified Borel transform will be left for future
investigations.

The inverse modified Borel transformation, corresponding to eq.~\eqn{mBTMJ},
is given by
\begin{equation}
\label{ImBTMJ}
\wt C_\um(\ah_Q) \,\equiv\, \int\limits_0^\infty\!{\rm d}t\,
{\rm e}^{-t/\ah_Q}\, \biggl(\frac{t}{\ah_Q}\biggr)^{\lambda t}\,
\frac{{\cal B}[\wt C_\um](t)}{\Gamma(1+\lambda t)} \,.
\end{equation}
Obviously, for $\beta_2=\lambda=0$, like in the large-$\beta_0$ approximation,
this definition goes over into the conventional Borel transformation.
Integrating the Borel transform order by order and employing the interesting
integral relation
\begin{equation}
\label{IntRel2}
\int\limits_0^\infty \!{\rm d}x\, {\rm e}^{-x}\,
\frac{n\,x^{n-1+z x}}{\Gamma(n+1+z x)} \,=\, 1 \,,
\end{equation}
as expected, one arrives at the Adler-function series of eq.~\eqn{C1tilde}.
The relation \eqn{IntRel2} can be proved in analogy to the derivation of
appendix~C of ref.~\cite{byz92}, page 4730, where the expanded form of the
modified Borel transform is confirmed.

Without loss of generality, let us again assume an IR renormalon pole with
respect to the conventional Borel transformation of the form
\begin{equation}
\label{IRpole}
B[\wt C_{\um,p}^{\rm IR}](u) \,=\, \frac{d_p^{\rm IR}}{(p-u)^\gamma} \,=\,
\frac{d_p^{\rm IR}}{p^\gamma (1-t/R)^\gamma} \,,
\end{equation}
with $R=2p/\beta_1$. The sub-leading terms of eq.~\eqn{BR3PIR} can be obtained
from different values of $\gamma$. Following appendix~C of ref.~\cite{byz92},
the conventional and modified Borel transforms are related by
\begin{equation}
\label{BhatB}
{\cal B}[\wt C_{\um}](t) \,=\, \lambda t \!\int\limits_0^1\! {\rm d}s\,
s\,(1-s)^{\lambda t-1} B[\wt C_{\um}](t s) \,.
\end{equation}
Inserting the Ansatz \eqn{IRpole}, and employing Euler's integral
representation of the hypergeometric function, we obtain
\begin{equation}
\label{Bhat2F1}
{\cal B}[\wt C_{\um,p}^{\rm IR}](t) \,=\,
\frac{d_p^{\rm IR}}{p^\gamma\,(1+\lambda t)}\,
{}_2F_1(\gamma,2;2+\lambda t;t/R) \,.
\end{equation}

We intend to confirm this relation by Taylor expansion. Employing the expansion
\begin{equation}
\frac{1}{(1-t/R)^\gamma} \,=\, \frac{1}{\Gamma(\gamma)}\,
\sum\limits_{n=0}^\infty \,\frac{\Gamma(n+\gamma)}{n!}\,
\biggl(\frac{t}{R}\biggr)^{\!n} \,,
\end{equation}
and integrating the conventional Borel transform term by term, one arrives at
\begin{equation}
\label{RPexp}
\wt C_{\um,p}^{\rm IR}(\ah_Q)\,=\,\frac{d_p^{\rm IR}}{p^\gamma\,\Gamma(\gamma)}
\,\sum\limits_{n=0}^\infty\, \frac{\Gamma(n+\gamma)}{R^n}\, (\ah_Q)^{n+1} \,,
\end{equation}
which coincides with the leading term of eq.~\eqn{Ral3PIR}. Extracting the
perturbative coefficients $\wt C_{\um,p}^{{\rm IR}(n)}$ from this expansion,
one obtains
\begin{equation}
\wt C_{\um,p}^{{\rm IR}(n)} \,=\, \frac{d_p^{\rm IR}}{p^\gamma} \,
\frac{\Gamma(n-1+\gamma)}{\Gamma(\gamma) R^{n-1}} \,.
\end{equation}
Now plugging these coefficients into the definition \eqn{mBTMJ} of the modified
Borel transform, and employing the series representation of the hypergeometric
function, one again finds the result \eqn{Bhat2F1}, which corroborates this
expression.

\begin{boldmath}
\subsection{Vector and scalar correlators in large-$\beta_0$}\label{sect4.3}
\end{boldmath}
\noindent
For a further discussion of the general structure of the Borel transform,
we briefly digress to the large-$\beta_0$ approximation, in which the Borel
transforms for Adler function and scalar correlator are known in closed form
to all orders in perturbation theory. Furthermore, in large-$\beta_0$ we have
the particular situation that the scheme dependence of the coupling only
originates from the fermion loop, and hence for each renormalisation scheme
RS a constant $\wh C$ exists, such that
\begin{equation}
\label{AQlb0}
\frac{1}{A_Q^{\beta_0}} \,\equiv\, \frac{1}{\ah_Q^{\wh C}} \,\equiv\,
\frac{1}{a_Q^{\rm{RS}}} + \frac{\beta_1}{2}\,{\wh C} \,=\,
\frac{1}{a_Q^{\MSb}} - \frac{5}{3}\,\frac{\beta_1}{2}
\end{equation}
is invariant \cite{ben98,jm16}. $A_Q^{\beta_0}$ can therefore be considered
a scheme-independent coupling in the large-$\beta_0$ approximation.

The conventional Borel transform \eqn{BC0t} for the reduced Adler function
of eq.~\eqn{Dhat} with respect to $A_Q^{\beta_0}$,
\begin{equation}
\hat D_{\beta_0}(Q^2) \,\equiv\, \frac{2}{\beta_1}\!\int\limits_0^\infty \!
{\rm d}u\, {\rm e}^{-2u/(\beta_1 A_Q^{\beta_0})} \hat B[\hat D_{\beta_0}](u)\,,
\end{equation}
has been calculated in refs.~\cite{ben92,bro92}, with the finding
\begin{equation}
\label{BDlb0}
\hat B[\hat D_{\beta_0}](u) \,=\, \frac{8\,C_F}{(2-u)}\,
\sum\limits_{k=2}^\infty\,\frac{(-1)^k k}{[k^2-(1-u)^2]^2} \,.
\end{equation}
Obviously, as both the correlator $\hat D_{\beta_0}(Q^2)$, as well as the
coupling $A_Q^{\beta_0}$, are independent of the scheme, also the Borel
transform $\hat B[\hat D_{\beta_0}](u)$ has to be scheme invariant, which is
reflected in the explicit expression \eqn{BDlb0}, and denoted by the hat on $B$.

Matters are a little more complicated for the scalar correlation function.
For this case, the large-$\beta_0$ approximation has been investigated in
detail in ref.~\cite{bkm00}. From these results, the Borel transform was then
extracted explicitly in ref.~\cite{jm16}. As can be observed from eq.~(3.16)
of \cite{jm16}, besides the Borel integral an additional logarithmic term is
present which depends on the scheme of the global coupling prefactor. The
structure of this term, however, is precisely such that it gets cancelled when
the prefactor is expressed in terms of the invariant coupling $A_Q^{\beta_0}$.
At leading order in large-$\beta_0$ the physical scalar correlation function
$\Psi^{''}(Q^2)$ can therefore be written in the following compact form:
\begin{equation}
\label{Psipplb0}
\Psi_{\beta_0}^{''}(Q^2) \,=\, \frac{N_c}{8\pi^2}\,
\frac{\hat m^2}{Q^2} \,\big( \pi A_Q^{\beta_0} \big)^{2\gamma_m^{(1)}/\beta_1}
\biggl\{\, 1 + \frac{2}{\beta_1}\! \int\limits_0^\infty \!{\rm d}u\,
{\rm e}^{-2u/(\beta_1 A_Q^{\beta_0})} \hat B[\Psi_{\beta_0}^{''}](u)
\,\biggr\} \,.
\end{equation}
The corresponding Borel transform $\hat B[\Psi_{\beta_0}^{''}](u)$ was found
to be \cite{bkm00,jm16}
\begin{equation}
\label{BhatPsi}
\hat B[\Psi_{\beta_0}^{''}](u) \,=\,
\frac{3}{2}\,C_F\,\Big[\, (1-u)\,G_D(u) - 1 \,\Big] \,,
\end{equation}
with the function $G_D(u)$ given by
\begin{equation}
G_D(u) \,=\, \frac{2}{1-u} - \frac{1}{2-u} +
\frac{2}{3} \sum\limits_{p=3}^\infty \frac{(-1)^p}{(p-u)^2} -
\frac{2}{3} \sum\limits_{p=1}^\infty \frac{(-1)^p}{(p+u)^2} \,,
\end{equation}
explicitly displaying the infrared and ultraviolet renormalon poles at
$u=2,\,3,\,4,\ldots$ and $u=-1,\,-2,\,-3,\ldots$ respectively, in analogy to
the case of the Adler function.

As is observed from eq.~\eqn{Psipplb0} above, the physical scalar current
correlation function in the large-$\beta_0$ approximation,
$\Psi_{\beta_0}^{''}(Q^2)$, can only be written in the form proportional to
$[\,1+\textrm{``Borel integral''}\,]$, if the global coupling prefactor is
expressed in terms of the invariant coupling $A_Q^{\beta_0}$. If this would
not be done, besides the Borel integral, additional polynomial terms related to
the anomalous dimensions of the involved currents would arise. As the reader
can convince oneself with the help of the results of ref.~\cite{jam12}, this
finding also holds true for the physical scalar gluonium correlator.

In full QCD, due to the substantially more complex structure, most probably
the concept of a universal scheme-invariant coupling does not exist. Hence,
generally the perturbative structure of a physical correlator, expressed in
terms of a Borel integral, reads
\begin{equation}
\label{InvStruct}
C_\um(\ah_Q) \,\sim\, [\ah_Q]^\delta \,\big[\, 1 + \textrm{``Borel integral''} +
\textrm{``polynomial terms''} \,\big] \,,
\end{equation}
where all involved components, coupling prefactor, Borel integral and
polynomial terms, depend on the renormalisation scheme of the QCD coupling.
Typically, as far as the perturbative expansion in the coupling is concerned,
the Borel integral will lead to an asymptotic series, while presumably the
polynomial contribution, even in full QCD, remains convergent.

As shall be discussed in more detail in the phenomenological section below,
depending on the employed scheme, the contribution of the polynomial terms
can be numerically significant, or even dominating over the contribution of
the Borel integral, at least for low orders, before the asymptotic behaviour,
governed by the renormalon singularities, sets in. This is for example the case
for the scalar correlator in the $\MSb$-scheme or the $C$-scheme at $C=0$. The
task therefore will be to identify schemes for which the polynomial terms are
small and the Borel integral provides the dominant contribution already for
lower orders. It will turn out that the corresponding schemes lead to a QCD
coupling close to the invariant coupling in large-$\beta_0$.

\vspace{4mm}
\subsection{Gluon condensate contributions to correlation functions}
\label{sect4.4}

In order to deduce the general renormalon structure of a given correlation
function, we require the higher-dimensional operator corrections in the
framework of the OPE. As we have seen from eq.~\eqn{gammat}, an infrared
renormalon pole at location $u=p$ is related to operators of dimension $d=2p$.
The lowest-dimensional gauge-invariant operator arising in the OPE of the
correlators under consideration in this work is the gluon condensate. Therefore,
we shall discuss those contributions for Adler function and scalar correlator
in some detail.

As was also indicated in section~\ref{sect4.1}, it is convenient to work with
an RGI basis of operators. Neglecting contributions from the quark condensate
and quartic mass corrections which are irrelevant for the renormalon structure,
the gluon condensate
$\langle O_{G^2}\rangle\equiv\langle G_{\mu\nu}^a G^{\mu\nu\,a}\rangle$ can be
reexpressed in terms of the scale-invariant gluon condensate \cite{sc88}
\begin{equation}
\langle \wh O_{G^2}\rangle \,=\, \frac{\beta(a_Q)}{\beta_1 a_Q}\,
\langle O_{G^2}\rangle \,=\,
\biggl( 1 + \frac{\beta_2}{\beta_1}\,a_Q + \ldots \biggr)
\langle a_Q\, G_{\mu\nu}^a G^{\mu\nu\,a}\rangle + \ldots \,.
\end{equation}
Employing the next-to-leading order result of ref.~\cite{st90} (eq.~(4.7)) for
the gluon-condensate contribution to the Adler function, and rewriting the QCD
coupling in terms of the $C$-scheme coupling $\ah_Q$, the Wilson coefficient
$\wh C_{G^2}^D(Q^2)$ for the scale invariant gluon condensate is found to be
\begin{equation}
\label{ChG2D}
\wh C_{G^2}^D(Q^2) \,=\, \frac{T_F}{3}\,\biggl[\, 1 + \biggl(
\frac{C_A}{2} - \frac{C_F}{4} - \frac{\beta_2}{\beta_1} \biggr) \ah_Q +
\ldots \,\biggr] \,.
\end{equation}
To compare renormalon pole residues in the next section below, it will
be convenient to have available the Wilson coefficient function in the
normalisation where the leading-order perturbative coefficient is taken out
as a global factor, such that the OPE perturbatively starts with a ``1''.
Employing the parton model result
\begin{equation}
\label{C00D}
C_\um^{(0)D} \,=\, \frac{N_c}{12\pi^2} \,,
\end{equation}
we finally obtain
\begin{equation}
\label{ChG2Dred}
\frac{\wh C_{G^2}^D(Q^2)}{C_\um^{(0)D}} \,=\, \frac{2\pi^2}{N_c}\,\biggl[\, 1 +
\biggl( \frac{C_A}{2} - \frac{C_F}{4} - \frac{\beta_2}{\beta_1} \biggr) \ah_Q +
\ldots \,\biggr] \,,
\end{equation}
where $T_F=1/2$ has been used.

Likewise, from the result of ref.~\cite{st90} for the scalar correlator
(eq.~(5.7)), and rewriting the QCD coupling in the global prefactor resulting
from the running quark mass once again in terms of the coupling $\ah_Q$, the
Wilson coefficient $\wh C_{G^2}^{\,\Psi^{''}}(Q^2)$ reads
\begin{equation}
\label{ChG2Psi}
\wh C_{G^2}^{\,\Psi^{''}}(Q^2) \,=\, \frac{T_F}{2}\,\frac{\wh m^2}{Q^2}\,
\big[\hat\alpha_s(Q)\big]^{2\gamma_m^{(1)}/\beta_1}\,\Big[\, 1 + 
\wt C_{G^2}^{(1)\Psi^{''}} \!\ah_Q + \ldots \,\Big] \,,
\end{equation}
with
\begin{equation}
\wt C_{G^2}^{(1)\Psi^{''}} \,=\,
\frac{3}{2}\,C_A + \frac{3}{4}\,C_F - \frac{\beta_2}{\beta_1} +
\biggl(\frac{3}{2} + C \biggr)\gamma_m^{(1)} + 2 \biggl(\frac{\gamma_m^{(2)}}{\beta_1} -
\frac{\beta_2 \gamma_m^{(1)}}{\beta_1^2}\biggr) \,.
\end{equation}
In this case, the ratio with the leading order perturbative result turns out
to be
\begin{equation}
\label{ChG2Psirat}
\frac{\wh C_{G^2}^{\,\Psi^{''}}(Q^2)}{C_\um^{(0)\Psi^{''}}} \,=\,
\frac{2\pi^2}{N_c}\,\Big[\, 1 + \wt C_{G^2}^{(1)\Psi^{''}} \!\ah_Q +
\ldots \,\Big] \,.
\end{equation}
It is observed that with $2\pi^2/N_c$ at leading order, this ratio is
the same for Adler function, eq.~\eqn{ChG2Dred}, and scalar correlator,
eq.~\eqn{ChG2Psirat}. As will be explained in the following section, this fact
will imply that also the corresponding renormalon pole residues are identical.

\subsection{Relation between renormalon-pole residues}\label{sect4.5}

In section~\ref{sect4.1} above, we had assumed that an inherent ambiguity
in the definition of the operators exists, such that they cancel against
corresponding ambiguities in the resummation of the perturbative series. This
lead to the relations of eq.~\eqn{gammat}. Now we intend to investigate the
cancellation of ambiguities a little further.

Employing the principal-value prescription for the integration over the
renormalon singularities in the Borel sum, the ambiguity on the perturbative
side should be purely imaginary. Without loss of generality, we can hence take
the ambiguity in the operator $\wh O_d$ to be of the form
$\pm\,i\Delta_p^{\rm IR} (\Lambda^{\!C})^d$. To ensure cancellation of the two
ambiguities, besides \eqn{gammat}, we must then have the relation
\begin{equation}
C_{O_d}^{(0)} \Delta_p^{\rm IR} \,=\, C_\um^{(0)} \,\biggl(\frac{2}{\beta_1}
\biggr)^{\!\gamma} \sin(\pi\gamma)\,\Gamma(1-\gamma)\, d_p^{\rm IR} \,,
\end{equation}
or equivalently
\begin{equation}
\biggl(\frac{\beta_1}{2}\biggr)^{\!\gamma}\,
\frac{\Delta_p^{\rm IR}}{\sin(\pi\gamma)\,\Gamma(1-\gamma)} \,=\,
\frac{C_\um^{(0)}}{C_{O_d}^{(0)}}\, d_p^{\rm IR} \,.
\end{equation}
The left-hand side of the last equation only depends on the operator considered.
Thus it should be universal, that is, independent of the correlation function
under investigation. Comparing two different correlators $D_A(Q)$ and $D_B(Q)$,
we hence find
\begin{equation}
\label{ResInv}
\frac{C_\um^{(0)}(A)}{C_{O_d}^{(0)}(A)}\, d_p^{\rm IR}(A) \,=\,
\frac{C_\um^{(0)}(B)}{C_{O_d}^{(0)}(B)}\, d_p^{\rm IR}(B) \,.
\end{equation}

The relation \eqn{ResInv} can be tested for different correlators that are
available in the large-$\beta_0$ approximation. Let us begin with the Adler
function. Employing the normalisation of eq.~\eqn{Pi10exp}, the perturbative
leading-order coefficient $C_\um^{(0)}$, as well as the one for the gluon
condensate $\wh C_{GG}^{(0)}$, take the values
\begin{equation}
\label{C0Adler}
C_\um^{(0)D} \,=\, \frac{N_c}{12\pi^2} \,, \qquad {\rm and} \qquad
C_{GG}^{(0)D} \,=\, \frac{1}{6} \,.
\end{equation}
The residue for the gluon-condensate renormalon pole at $u=2$ for the scheme
invariant Borel transform \eqn{BDlb0}, on the other hand, is given by
\begin{equation}
\label{d2Adler}
d_2^{\rm IR} \,=\, \frac{3}{2}\,C_F \,.
\end{equation}
Hence, for the case of the Adler function in large-$\beta_0$, the total
combination \eqn{ResInv} for the gluon-condensate renormalon pole reads:
\begin{equation}
\label{ResInvAdler}
\frac{C_\um^{(0)D}}{C_{GG}^{(0)D}}\, d_2^{\rm IR} \,=\,
\frac{3}{8\pi^2}\,(N_c^2-1) \,.
\end{equation}
For the scalar correlation function, the ratio of leading-order coefficients
was already calculated in eq.~\eqn{ChG2Psirat} in the last sub-section, with
the result
\begin{equation}
\frac{C_\um^{(0)\Psi^{''}}}{C_{G^2}^{(0)\Psi^{''}}} \,=\,
\frac{N_c}{2\pi^2} \,,
\end{equation}
identical to the one for the Adler function. for this reason the residue for
the renormalon pole at $u=2$ should coincide with eq.~\eqn{d2Adler}. Inspection
of eq.~\eqn{BhatPsi} confirms that this is indeed the case. As an additional
test, one can check that the relation \eqn{ResInvAdler} is also satisfied by
the corresponding results for the scalar gluonium correlation function in the
large-$\beta_0$ approximation \cite{jam12}.

\section{Borel models}\label{sect5}

The aim in this section is to construct models for the Borel transform of the
two correlation functions under investigation, along the lines of the work
of ref.~\cite{bj08} for the Adler function, but employing the $C$-scheme
coupling $\ah_Q$.

Let us begin with outlining the general philosophy of the Borel models. At
large orders, the perturbative series for both correlators will be dominated
by the renormalon lying closest to $u=0$. In the cases at hand this is the
leading UV renormalon with a pole at $u=-1$. At intermediate orders, low lying
IR renormalons, the lowest one being at $u=2$ related to the gluon condensate,
and at $u=3$ connected with dimension-6 operators, should provide significant
contributions. Finally, at the very lowest orders, in general, renormalon
dominance of only a few poles cannot yet be expected to be realised.

The relative importance of different renormalon contributions, however, also
depends on the renormalisation scheme. Since the structure of the $u=2$ IR
renormalon is known best, it is desirable to work in a scheme where the last
analytically known perturbative orders, typically the third and fourth, receive
a sizeable contribution from this and perhaps the next renormalon. Below it
shall be argued that this is achieved in the $\MSb$ scheme, or the $C$-scheme
with $C=0$, which is close to $\MSb$. For the scalar correlator, on the other
hand, as was discussed in section \ref{sect4.3}, a scheme close to the invariant
one in large-$\beta_0$ should be chosen for the global $\as$-prefactor, in
order to remove large polynomial contributions related to anomalous dimensions.

\subsection{Borel model for the Adler function}\label{sect5.1}

Regarding the Borel model for the Adler function, we closely follow the
work of ref.~\cite{bj08}, and the reader is referred to this article, in
particular section~6, for further details. The model for the Borel transform
of $\wt C_{\um}(\ah_Q)$ that we advocate here takes the form
\begin{equation}
\label{BRu}
B[\wt C_{\um}](u) \,=\, B[\wt C_{\um,2}^{\rm IR}](u) +
B[\wt C_{\um,3}^{\rm IR}](u) + B[\wt C_{\um,1}^{\rm UV}](u) +
d_0^{\rm PO} \,.
\end{equation}
The general structure of an IR renormalon pole at position $p$ was provided
in eq.~\eqn{BR3PIR}. In the spirit of ref.~\cite{bj08}, we have included the
lowest lying IR renormalon pole at $u=2$, for which most information is
available, as well as the next-to-leading IR pole at $u=3$, together with
a UV pole at $u=-1$ which dominates the perturbative series at large orders.
Regarding the polynomial contribution which should take care of very low
orders, we have only included the constant $d_0^{\rm PO}$. In ref.~\cite{bj08}
it was found that this is sufficient to obtain compatible models, and it has
the advantage that all unknown parameters, the three residues of the renormalon
poles and the constant $d_0^{\rm PO}$ can be adjusted such that the four
analytically known perturbative coefficients $\bar c_{1,1}$ to $\bar c_{4,1}$
of eqs.~\eqn{cn1} and \eqn{cb3cb4} are reproduced.

Let us discuss the explicit structure of the included renormalon poles in some
more detail. The OPE term corresponding to the gluon condensate was discussed
in section~\ref{sect4.4}, and the Wilson coefficient for the Adler function
was provided in eq.~\eqn{ChG2D}. From this result we can deduce the parameters
required in eq.~\eqn{gammat}, namely $\eta=0$ and the next-to-leading order
correction $b_1$, hence fixing the $u=2$ renormalon pole up to NLO apart from
the residue $d_2^{\rm IR}$. For the IR pole at $u=3$, matters are substantially
more complicated because several dimension-6 operators, four-quark condensates
and the triple-gluon condensate, contribute. A detailed discussion of the
dimension-6 NLO Wilson coefficient for the Adler function was given in
ref.~\cite{bhj15}. As it is impossible to include several poles with the
corresponding anomalous dimensions, due to the many additional unknown residue
parameters, we have decided to only include the strongest pole, that is, the
one with the highest power in the exponent. For the dimension-6 four-quark
operators, the contribution $\eta=1$ is largely cancelled by the
anomalous-dimension term, such that with $\gamma\approx 3.1$ the exponent
is close to the one already employed in ref.~\cite{bj08}. Still, it is much
stronger than in the large-$\beta_0$ case where at most quadratic renormalon
divergences are found. Since the $u=3$ pole is only included in an effective
sense, we have not incorporated a NLO correction, even though they are
available.  Finally, the general structure of a UV renormalon pole was provided
in ref.~\cite{ben98} and section~5 of ref.~\cite{bj08}. Compared to the latter
work we have also included the contribution of the anomalous dimension which
yields the strongest pole, leading to $\gamma\approx 1.6$. However, because up
fourth order the leading UV renormalon only has very little influence on the
perturbative coefficients, the changes in our numerics turn out to be minor.

The last issue before being able to apply our model numerically is the question
which renormalisation scheme is most adequate. In the $C$-scheme, this can
be easily investigated by tuning the scheme-parameter $C$. Performing this
exercise, it is found that around $C\approx -0.7$ the second order coefficient
$\hat c_{2,1}$ turns negative as the first of the analytically known
coefficients. This indicates that at such and even more negative $C$, dominance
of the leading UV renormalon sets in at rather low orders, because only the UV
renormalons contribute in a sign-alternating fashion. On the other hand, if $C$
is chosen positive and too large, the lowest IR renormalon at $u=2$ does not
yet dominate at the last known perturbative orders. We shall advocate here that
the choice $C=0$, being close to the $\MSb$ scheme, appears most adequate. This
will also be corroborated by our numerical analysis to which we turn next.

Adjusting the four unknown parameters of our model, the three renormalon-pole
residues $d_2^{\rm IR}$, $d_3^{\rm IR}$ and $d_1^{\rm UV}$, as well as the
constant $d_0^{\rm PO}$, to the four known perturbative coefficients
$\bar c_{1,1}$ to $\bar c_{4,1}$ of eqs.~\eqn{cn1} and \eqn{cb3cb4}, we obtain
\begin{equation}
\label{dIRUV}
d_2^{\rm IR} \,=\,    2.67  \,, \qquad
d_3^{\rm IR} \,=\, -\,7.23 \,, \nn \\[1mm]
\end{equation}
\begin{equation}
d_1^{\rm UV} \,=\, -\,1.35\cdot 10^{-2} \,, \qquad
d_0^{\rm PO} \,=\,    0.273 \,.
\end{equation}
The values of these four parameters turn out similar to the results of
ref.~\cite{bj08}. The Borel model now allows to predict still higher-order
coefficients. To be conservative, in this work we restrict ourselves to only
quote the next unknown, fifth order coefficient which assumes the value
$\bar c_{5,1}=310.4$. The $C$-scheme result at $C=0$ can be converted to an
$\MSb$-scheme value with the finding $c_{5,1}=245.0$. This $c_{5,1}$ turns out
only about 15\% smaller than the corresponding coefficient advocated in
ref.~\cite{bj08}, which might give an idea of the uncertainties induced by
reasonable assumptions about the Borel model. Our result can be compared to
other recent predictions for $c_{5,1}$ from Pad\'e approximants \cite{bmo18},
$c_{5,1}=277\pm 51$, or conformal mappings \cite{cap19}, $c_{5,1}=287\pm 40$,
finding good agreement. Still, to be substantially more conservative, in
phenomenological applications the obtained $c_{5,1}$ might be employed with
100\% uncertainty as an estimate for yet unaccounted higher orders.

To conclude this sub-section, we investigate how large the contribution of
the three renormalon poles is to a given perturbative coefficient. Taking the
last included fourth order coefficient $\bar c_{4,1}$, the contributions of
the $u=2$ and $u=3$ IR poles, and the $u=-1$ UV pole, are respectively $122\%$,
$-34\%$ and $12\%$. This nicely confirms that the coefficient is dominated by
the lowest lying IR renormalon pole at $u=2$, with a tolerable contribution of
the next IR pole at $u=3$, and a still small correction of the leading UV pole
at $u=-1$, such that the hierarchy of contributions is realised as asserted
above. The last observation provides further confidence in the applicability
of models for the Borel transform of perturbative series as an estimator for
as yet unknown higher orders.

\subsection{Borel model for the scalar correlator}\label{sect5.2}

In section~4.3 above, it was already observed that the generic form of a
physical, RG invariant, two-point correlator, for which the initial current
has an anomalous dimension, like in the case of the scalar correlator, can
suggestively be expressed as:
\begin{equation}
\label{InvStruct2}
\Psi_{\rm PT}^{''}(Q^2) \,\sim\, \Big[\ah_Q^{C_m}\Big]^\delta \,\Big\{\, 1 +
\textrm{``Borel integral''}\!\left(\ah_Q^{C_a}\right) +
\textrm{``polynomial terms''}\!\Big(\ah_Q^{C_a}\Big) \,\Big\} \,.
\end{equation}
In writing eq.~\eqn{InvStruct2}, it is already indicated, that the global
factor, and the terms in curly brackets, can be expressed in $C$-scheme
couplings at different $C$, namely $C_m$ and $C_a$. The motivation behind
this choice is as follows: if we aim at fitting a Borel model for the scalar
correlator, we only have at our disposal the first four known perturbative
coefficients. However, the central Borel model that was advocated for the
Adler function contains four free parameters. Therefore, it is impossible to
extract additional information on the polynomial contribution, and we have to
try to make it as small as possible. This can be achieved by tuning the global
scheme parameter $C_m$. In the large-$\beta_0$ approximation, for a particular
choice of $C$, it was possible to make the polynomial contribution vanish,
and then the global factor was expressed in terms of the invariant coupling
$A_Q^{\beta_0}$. As presumably no universal, invariant coupling exists in full
QCD, we can not hope to completely remove the polynomial contribution by
varying $C_m$.\footnote{This can also be inferred from the general form of an
invariant, polynomial contribution in the $C$-scheme, which is presented in
appendix~B.} Nonetheless, it should be possible to make it substantially
smaller than the contribution from the Borel integral, such that we have a
chance to extract a reasonable Borel model, also for the scalar correlation
function.

In addition, in section~4.5 we have seen that the combination \eqn{ResInv}
of Wilson coefficient functions and renormalon-pole residues is universal.
Furthermore, the ratio of Wilson coefficient normalisations was even found
identical for Adler function and scalar correlator in the case of the gluon
condensate contribution. Hence, the residue of the gluon condensate renormalon
pole is no longer a free parameter, but can be taken from the Borel model for
the Adler function. This allows to introduce one additional free parameter
which can be fixed from the four available perturbative coefficients. The idea
now is to employ $C_m$ as this parameter, and to fix it together with the
remaining three parameters of the Borel model, namely $d_3^{\rm IR}$,
$d_1^{\rm UV}$ and $d_0^{\rm PO}$. In this way, one might get an idea which
range of scales is most compatible with a domination of the Borel integral
contribution, and a negligible polynomial term.

Proceeding in this fashion, the central Borel model for the Adler function
is also assumed for the Borel contribution of the scalar correlation function.
It only remains to replace the NLO contribution for the gluon-condensate
renormalon pole according to the results of the Wilson coefficient function
of section~4.4, and to fix a scheme choice for the coupling in the Borel
integral. For simplicity, here $C_a=0$ is taken, in analogy to the analysis
for the Adler function. Adjusting all parameters such that the perturbative
coefficients of eq.~\eqn{rb1torb4} are reproduced, one obtains:
\begin{equation}
\label{dIRUVscalar}
d_2^{\rm IR} \,=\,    2.67  \,, \qquad
d_3^{\rm IR} \,=\, -\,43.69 \,, \qquad
d_1^{\rm UV} \,=\, -\,4.68\cdot 10^{-2} \,, \nn \\[1mm]
\end{equation}
\begin{equation}
d_0^{\rm PO} \,=\,    0.400 \,, \qquad
C_m \,=\, -\,1.603 \,.
\end{equation}
Also inspecting the coefficients of the perturbative expansion within the
curly brackets of eq.~\eqn{InvStruct2}, one finds
\begin{equation}
\label{CurlyBrack}
\Psi_{\rm PT}^{''}(Q^2) \,\sim\, \Big[\ah_Q^{C_m}\Big]^\delta \Big\{\, 1 +
2.25\,\ab_Q + 1.62\,\ab_Q^2 + 2.72\,\ab_Q^3 + 72.4\,\ab_Q^4 + 155\,\ab_Q^5 +
\ldots \,\Big\} \,.
\end{equation}
Several observations can be made on the basis of these results. First of all
it is interesting to note that the preferred value of $C_m$ is rather close
to $C_m=-5/3$, in large-$\beta_0$ leading to the invariant coupling
$A_Q^{\beta_0}$. This suggests that a large fraction of the perturbative
correction to the scalar correlator in the $\MSb$ scheme is due to the
polynomial contribution, and can be resummed into the prefactor through an
appropriate scheme choice. Then, the behaviour of the remaining perturbative
corrections, that dominantly originate from the Borel integral, are much more
``Adler function''-like.

Regarding the parameters of the Borel model, the constant $d_0^{\rm PO}$ turns
out small which indicates that even the lowest order correction already
receives a substantial contribution from the renormalon poles. The residue
of the $u=3$ term is found very different from the Adler function.  On the
one hand, however, the dimension-6 contributions in the scalar correlator are
different from the Adler function, and furthermore this pole also effectively
parametrises all missing IR renormalon poles that are not included in the model.
Therefore, no conclusions can be drawn on the basis of this residue. Finally,
the residue of the leading UV pole, $d_1^{\rm UV}$ is found larger than the
one for the Adler function, which entails that the scalar correlator reaches
its asymptotic behaviour at lower orders than the Adler function. In fact, 
investigating even higher orders to eq.~\eqn{CurlyBrack}, it is found that
the 7th order term is negative, while for the Adler function series in the
$C=0$ scheme the first negative coefficient is found at the 9th order.

As a prediction of the model, we are now in a position to compute the fifth
order coefficient $d_{5,1}^{\,''}$ to the scalar correlation function in
the $\MSb$ scheme, which supplements eq.~\eqn{dt11todt41n}. Performing all
necessary scheme transformations, one arrives at
\begin{equation}
\label{d51ppmsb}
d_{5,1}^{\,''} \,=\, 3201 - 9.88\cdot 10^{-3}\beta_6 -
                            8.89\cdot 10^{-2}\gamma_6 \,,
\end{equation}
% {d5pp -> 3200.82 - 0.00987654 b6 - 0.0888889 g6}
% {-1., -15.8333, -194.239, -2538.44, -36913.5}  %  d_n,2
which, however, also depends on the as yet unknown RG coefficients $\beta_6$
and $\gamma_6$. Therefore, in order to make definite predictions for the
perturbative correction of the scalar correlator at the fifth order, the
6th order RG coefficients are required as well. Also making use of the dependent
coefficient $d_{5,2}=-\,36913.5$, which can be calculated from eq.~\eqn{dn2},
one finally finds the Borel model expectation
\begin{equation}
\label{d51msb}
d_{5,1} \,=\, 77028 - 9.88\cdot 10^{-3}\beta_6 -
                      8.89\cdot 10^{-2}\gamma_6 \,.
\end{equation}
Hence, we see that the $\MSb$ coefficient $d_{5,1}$ is largely dominated by
the contribution from $d_{5,2}$. A more detailed investigation of Borel models
for the scalar correlator, and other two-point correlation functions based
on currents with non-vanishing anomalous dimension, will be relegated to a
forthcoming publication in the future \cite{jm36}.

\section{Summary}\label{sect6}

Information on the higher-order behaviour of perturbative series is important 
in order to obtain reliable error estimates for phenomenological predictions.
As the perturbative expansions in full QCD most probably only lead to asymptotic
series, its Borel transforms, which have convergent expansions, are interesting
objects to be studied. Making use of constraints originating from the operator
product expansion and the renormalisation group, simple models for the Borel
transforms of two-point correlation functions in QCD can be written down.

In this work, a Borel model for the vector correlation function, or Adler
function, that has already been presented in the literature \cite{bj08}, was
reviewed. A novel ingredient here is that the Borel model has been expressed
in terms of the so-called $C$-scheme coupling \cite{bjm16}. The $C$-scheme
coupling has several advantages: on the one hand, the $\beta$-function of the
$C$-scheme coupling only depends on the invariant coefficients $\beta_1$ and
$\beta_2$, and assumes a simple geometric form. Hence it is known to all
orders in perturbation theory. For this reason, the general form of a certain
renormalon pole in the $C$-scheme turns out simpler than for example in the
$\MSb$ scheme. Finally, variations of the renormalisation scheme are easily
realised through changes in the scheme parameter $C$.

On the phenomenological side, the Borel model allows to predict yet unknown
higher-order coefficients in the perturbative series. To be conservative, in
this article only an estimate of the next, fifth-order coefficient has been
presented. For the Adler function this was discussed in detail in section
\ref{sect5.1}, with the result $\bar c_{5,1}=310.4$ in the $C$-scheme.
Transforming this value back into the $\MSb$ scheme, one obtains
$c_{5,1}=245.0$, which may be confronted by future analytical computations.
It turns out similar to the results already obtained in
refs.~\cite{bj08,bmo18,cap19}~directly in the $\MSb$ scheme, thereby
providing support to these analyses.

As a second two-point correlator, the scalar correlation function was
investigated, and a Borel model for its perturbative series has been presented
in section~\ref{sect5.2}. The corresponding construction is new and has as yet
not been presented elsewhere. Due to the fact that the scalar current carries
an anomalous dimension, its structure is substantially more complicated. It
turns out that the global factor of $\alpha_s$ entails that, besides the Borel
integral, additional contributions are present that depend on the anomalous
dimension of the initial current. Depending on the scheme choice, these
contributions can be large, even dominating the series, which seems to be the
case for the scalar correlator in the $\MSb$ scheme. Extracting a Borel model
necessitates that the additional, polynomial contribution is largely removed
by a scheme change. Interestingly enough, the required scheme turns out close
to the invariant coupling in large-$\beta_0$. On the phenomenological side,
in eq.~\eqn{d51msb} an estimate was also given for the unknown, fifth-order
coefficient $d_{5,1}$ of the scalar correlator. This prediction, however,
depends on the so far uncalculated RG coefficients $\beta_6$ and $\gamma_6$,
so that it remains unclear, when it might be tested against a fully analytical
calculation of the fifth perturbative order for the scalar correlator.

\section*{Acknowledgments}
Partial collaboration in this work with Ramon~Miravitllas, and interesting
discussions with Andre~Hoang, are gratefully acknowledged. The author would
also like to thank the FWF Austrian Science Fund under the Project
No.~P28535-N27 for partial support, and the particle physics group at the
University of Vienna, where part of this work was completed. 
\vspace*{0.3cm}

\appendix
\section{Renormalisation group functions and dependent coefficients}\label{appA}

In our notation, the QCD $\beta$-function and mass anomalous dimension are
defined as:
\begin{eqnarray}
\label{bega}
-\,\mu\,\frac{{\rm d}a}{{\rm d}\mu} &\equiv& \beta(a) \,=\,
\beta_1\,a^2 + \beta_2\,a^3 + \beta_3\,a^4 + \beta_4\,a^5 + \ldots \,, \\
\tvs
-\,\frac{\mu}{m}\,\frac{{\rm d}m}{{\rm d}\mu} &\equiv& \gamma_m(a) \,=\,
\gamma_m^{(1)}\,a + \gamma_m^{(2)}\,a^2 + \gamma_m^{(3)}\,a^3 +
\gamma_m^{(4)}\,a^4 + \ldots \,.
\end{eqnarray}
It is assumed that we work in a mass-independent renormalisation scheme
and in this study throughout the modified minimal subtraction scheme $\MSb$
is used. To make the presentation self-contained, below the known coefficients
of the $\beta$-function and mass anomalous dimension in the given conventions
shall be provided. Numerically, for $N_c=3$ and $N_f=3$, the first five
coefficients of the $\beta$-function are given by \cite{tvz80,lrv97,cza04,bck16}
\begin{eqnarray}
\label{bfun}
\beta_1 &=& \sfrac{9}{2} \,, \qquad
\beta_2 \,=\, 8 \,, \qquad
\beta_3 \,=\, \sfrac{3863}{192} \,, \qquad
\beta_4 \,=\, \sfrac{140599}{2304} + \sfrac{445}{16}\,\zeta_3 \,, \nn \\
\tvs
\beta_5 &=& \sfrac{139857733}{663552} + \sfrac{11059213}{27648}\,\zeta_3 -
\sfrac{36045}{512}\,\zeta_4 - \sfrac{534385}{1536}\,\zeta_5 \,,
\end{eqnarray}
and the first five for $\gamma_m(a)$ are found to be \cite{vlr97,bck14}
\begin{eqnarray}
\label{gfun}
\gamma_m^{(1)} &=& 2 \,, \qquad
\gamma_m^{(2)} \,=\, \sfrac{91}{12} \,, \qquad
\gamma_m^{(3)} \,=\, \sfrac{8885}{288} - 5\,\zeta_3 \,, \nn \\
\tvs
\gamma_m^{(4)} &=& \sfrac{2977517}{20736} - \sfrac{9295}{216}\,\zeta_3 +
\sfrac{135}{8}\,\zeta_4 - \sfrac{125}{6}\,\zeta_5 \,, \\
\tvs
\gamma_m^{(5)} &=& \sfrac{156509815}{248832} - \sfrac{23663747}{62208}\,\zeta_3
+ 170\,\zeta_3^2 + \sfrac{23765}{128}\,\zeta_4 -
\sfrac{22625465}{31104}\,\zeta_5 + \sfrac{1875}{16}\,\zeta_6 +
\sfrac{118405}{288}\,\zeta_7 \nn \,.
\end{eqnarray}

The dependent perturbative coefficients $d_{n,k}$ with $k>1$ can be expressed
in terms of the independent coefficients $d_{n,1}$, and coefficients of the
QCD $\beta$-function and mass anomalous dimension. In particular, the
coefficients $d_{n,2}$, which are required in eq.~\eqn{Psippres}, take the form
\begin{equation}
\label{dn2}
d_{n,2} \,=\, -\,\frac{1}{2}\,\gamma_m^{(n)} d_{0,1} - \frac{1}{4}
\sum\limits_{k=1}^{n-1} \big( 2\gamma_m^{(n-k)} + k\,\beta_{n-k} \big)
d_{k,1} \,.
\end{equation}

\section{General scheme-invariant structure}\label{appB}

In this appendix, the general scheme-invariant structure of a two-point
correlation function in the $C$-scheme will be provided, which for example
has to be obeyed by the polynomial contribution in eq.~\eqn{InvStruct}.
Denoting the structure by $P(\ah_Q)$, it takes the general form
\begin{equation}
P(\ah_Q) \,=\, [\ah_Q]^\delta \,\biggl\{\, 1 + \sum\limits_{n=1}^\infty
(\ah_Q)^n \sum\limits_{k=0}^n y_{n,k}\, {\wh C}^{\,k} \,\biggr\} \,,
\end{equation}
where $\wh C\equiv\beta_1/2\,C$. Relations between the coefficients $y_{n,k}$
can be obtained from the RG equation~\eqn{CRGE}. Up to order $\ah_Q^4$, and
setting $\lambda = \beta_2/\beta_1$, those relations read:
\begin{eqnarray}
y_{1,1} &=& \delta \,, \qquad
y_{2,2} \,=\, \frac{\delta}{2} \,(\delta+1) \,, \qquad
y_{2,1} \,=\, (\delta+1)\,y_{1,0} + \lambda\,\delta \,, \nn \\
\mvs
y_{3,3} &=& \frac{\delta}{6}\,(\delta+1)(\delta+2) \,, \qquad
y_{3,2} \,=\, \frac{1}{2}\,\big[ (\delta+1)(\delta+2)\,y_{1,0} +
\lambda\,\delta\,(2\delta+3) \big] \,, \nn \\
\mvs
y_{3,1} &=& (\delta+2)\,y_{2,0} + \lambda\,(\delta+1) \,y_{1,0} +
\lambda^2 \delta \,, \qquad
y_{4,4} \,=\, \frac{\delta}{24}\,(\delta+1)(\delta+2)(\delta+3) \,, \nn \\
\mvs
y_{4,3} &=& \frac{1}{6}\,\big[ (\delta+1)(\delta+2)(\delta+3)\,y_{1,0} +
\lambda\,\delta\,(3\delta^2+12\delta+11) \big] \,, \\
\mvs
y_{4,2} &=& \frac{1}{6}\,\big[ (\delta+2)(\delta+3)\,y_{2,0} + \lambda\,
(\delta+1)(2\delta+5) \,y_{1,0} + 3\lambda^2\delta\,(\delta+2) \big] \,, \nn \\
\mvs
y_{4,1} &=& (\delta+3)\,y_{3,0} + \lambda\,(\delta+2) \,y_{2,0} +
\lambda^2 (\delta+1)\,y_{1,0} + \lambda^3\delta \,. \nn
\end{eqnarray}
If still higher orders are required, it is an easy matter to compute them from
the RG equation. Like for the two-point correlators, the coefficients $y_{n,0}$
cannot be determined from the renormalisation group, and can be considered
independent.


\begin{thebibliography}{}

\bibitem{gzj90}
J.C.~Le~Guillou and J.~Zinn-Justin,
{\em Large order behavior of perturbation theory}, North-Holland (1990).

\bibitem{ben98}
M.~Beneke, {\it Renormalons},  {\em Phys.\ Rept.} {\bf 317} (1999) 1--142,
[\href{http://arxiv.org/abs/hep-ph/9807443}{{\tt hep-ph/9807443}}].

\bibitem{adl74}
S.L. Adler,
{\it {Some simple vacuum polarization phenomenology: $e^+e^-\to\,$ Hadrons}},
{\em Phys. Rev.} {\bf D10} (1974) 3714.

\bibitem{bj08}
M.~Beneke and M.~Jamin,
{\it $\alpha_s$ and the $\tau$ hadronic width: fixed-order, contour-improved
 and higher-order perturbation theory},
{\em JHEP} {\bf 0809}   (2008) 044,
\href{http://arxiv.org/abs/0806.3156}{{\tt arXiv:0806.3156 [hep-ph]}}.

\bibitem{bbj12}
M.~Beneke and D.~Boito and M.~Jamin,
{\it Perturbative expansion of $\tau$ hadronic spectral function moments and $\alpha_s$ extractions},
{\em JHEP} {\bf 01}   (2013) 125,
\href{http://arxiv.org/abs/1210.8038}{{\tt arXiv:1210.8038 [hep-ph]}}.

\bibitem{bjm16}
D.~Boito, M.~Jamin, and R.~Miravitllas,
{\it Scheme variations of the {QCD} coupling and hadronic $\tau$ decays},
{\em Phys.\ Rev.\ Lett.} {\bf 117} (2016) 152001,
[\href{http://arxiv.org/abs/1606.06175}{{\tt arXiv:1606.06175 [hep-ph]}}].

\bibitem{jm16}
M.~Jamin and R.~Miravitllas,
{\it Scalar correlator, {Higgs} decay into quarks, and scheme variations of
 the {QCD} coupling},
{\em JHEP} {\bf 1610} (2016) 059,
[\href{http://arxiv.org/abs/1606.06166}{{\tt arXiv:1606.06166 [hep-ph]}}].

\bibitem{bbdm78}
W.A.~Bardeen, A.J.~Buras, D.W.~Duke and T.~Muta,
{\it Deep inelastic scattering beyond the leading order in asymptotically
 free gauge theories},
{\em Phys.\ Rev.\ D} {\bf 18} (1978) 3998.

\bibitem{wsdb18}
X.G.~Wu, J.M.~Shen, B.L.~Du and S.J.~Brodsky,
{\it Novel demonstration of the renormalization group invariance of the
 fixed-order predictions using the principle of maximum conformality and
 the $C$-scheme coupling},
{\em Phys.\ Rev.\ D} {\bf 97} (2018) 094030,
[\href{http://arxiv.org/abs/1802.09154}{{\tt arXiv:1802.09154 [hep-ph]}}].

\bibitem{gkl91}
S.G. Gorishnii, A.L. Kataev, and S.A. Larin,
{\it {The ${\cal O}(\alpha_s^3)$ corrections to
 $\sigma_{{\rm tot}}(e^+ e^- \to {\rm hadrons})$ and
 ${\Gamma}(\tau^- \to \nu_\tau + {\rm hadrons})$ in QCD}},
{\em Phys. Lett.} {\bf B259} (1991) 144.

\bibitem{ss91}
L.R. Surguladze and M.A. Samuel,
{\it {Total hadronic cross-section in $e^+ e^-$ annihilation at the four-loop
 level of perturbative QCD}},
{\em Phys. Rev. Lett.} {\bf 66} (1991) 560.

\bibitem{bck08}
P.A. Baikov, K.G. Chetyrkin, and J.H. K{\"u}hn,
{\it {Hadronic $Z$- and $\tau$-Decays in Order $\alpha_s^4$}},
{\em Phys. Rev. Lett.} {\bf 101} (2008) 012002,
\href{http://arxiv.org/abs/0801.1821}{{\tt arXiv:0801.1821 [hep-ph]}}.

\bibitem{bjm17}
D.~Boito, M.~Jamin, and R.~Miravitllas,
{\it Scheme variations of the {QCD} coupling},
{\em EPJ Web Conf.} {\bf 137} (2017) 05007,
[\href{http://arxiv.org/abs/1612.01792}{{\tt arXiv:1612.01792 [hep-ph]}}].

\bibitem{gkls90}
S.G.~Gorishnii, A.L.~Kataev, S.A.~Larin, and L.R.~Surguladze,
{\it Corrected three loop {QCD} correction to the correlator of the quark
 scalar currents and {$\Gamma_{\rm tot}(H_0\to{\rm hadrons})$}},
{\em Mod.\ Phys.\ Lett.\ A} {\bf 5} (1990) 2703.

\bibitem{che96}
K.G.~Chetyrkin,
{\it Correlator of the quark scalar currents and {$\Gamma_{\rm tot}(H_0\to
 {\rm hadrons})$} at {${\cal O}(\alpha_s^3)$} in p{QCD}},
{\em Phys.\ Lett.\ B} {\bf 390} (1997) 309,
[\href{http://arxiv.org/abs/hep-ph/9608318}{{\tt hep-ph/9608318}}].

\bibitem{bck05}
P.A.~Baikov, K.G.~Chetyrkin, and J.H.~K{\"u}hn,
{\it Scalar correlator at {${\cal O}(\alpha_s^4)$}, {Higgs} decay into
 $b$-quarks and bounds on the light quark masses},
{\em Phys.\ Rev.\ Lett.} {\bf 96} (2006) 012003,
[\href{http://arxiv.org/abs/hep-ph/0511063}{{\tt hep-ph/0511063}}].

\bibitem{bro81}
D.J.~Broadhurst,
{\it Chiral symmetry breaking and perturbative QCD},
{\em Phys.\ Lett.\ } {\bf 101B} (1981) 423.

\bibitem{bck17}
P.A.~Baikov, K.G.~Chetyrkin, and J.H.~K{\"u}hn,
{\it Five-loop fermion anomalous dimension for a general gauge group from
 four-loop massless propagators},
{\em JHEP} {\bf 1704} (2017) 119,
[\href{http://arxiv.org/abs/1702.01458}{{\tt arXiv:1702.01458 [hep-ph]}}].

\bibitem{jm17}
M.~Jamin and R.~Miravitllas,
{\it Absence of even-integer $\zeta$-function values in Euclidean physical
 quantities in QCD},
{\em Phys.\ Lett.\ B} {\bf 779} (2018) 452,
[\href{http://arxiv.org/abs/1711.00787}{{\tt arXiv:1711.00787 [hep-ph]}}].

\bibitem{dv17}
J.~Davies and A.~Vogt,
{\it Absence of $\pi^2$ terms in physical anomalous dimensions in DIS:
 Verification and resulting predictions},
{\em Phys.\ Lett.\ B} {\bf 776} (2018) 189,
[\href{http://arxiv.org/abs/1711.05267}{{\tt arXiv:1711.05267 [hep-ph]}}].

\bibitem{bc18}
P.A.~Baikov and K.G.~Chetyrkin,
{\it The structure of generic anomalous dimensions and no-$\pi$ theorem for
 massless propagators},
[\href{http://arxiv.org/abs/1804.10088}{{\tt arXiv:1804.10088 [hep-ph]}}].

\bibitem{bhj15}
D.~Boito, D.~Hornung and M.~Jamin,
{\it Anomalous dimensions of four-quark operators and renormalon structure of mesonic two-point correlators},
{\em JHEP} {\bf 1512} (2015) 090,
[\href{http://arxiv.org/abs/1510.03812}{{\tt arXiv:1510.03812 [hep-ph]}}].

\bibitem{jam12}
M.~Jamin,
{\it The scalar gluonium correlator: large-$\beta_0$ and beyond},
JHEP {\bf 1204} (2012) 099,
[\href{http://arxiv.org/abs/1202.1169}{{\tt arXiv:1202.1169 [hep-ph]}}].

\bibitem{byz92}
L.S.~Brown, L.G.~Yaffe and C.X.~Zhai,
{\it Large-order perturbation theory for the electromagnetic current current
 correlation function},
{\em Phys.\ Rev.\ D} {\bf 46} (1992) 4712,
[\href{http://arxiv.gov/abs/hep-ph/9205213}{\tt hep-ph/9205213}].

\bibitem{gru93}
G.~Grunberg,
{\it The renormalization scheme invariant Borel transform and the QED
 renormalons},
{\em Phys.\ Lett.\ B} {\bf 304} (1993) 183.

\bibitem{bo20}
D.~Boito, F.~Oliani,
{\it Renormalons in integrated spectral function moments and $\alpha_s$ extractions},
{\em Phys.\ Rev.\ D} {\bf 101} (2020) 074003,
[\href{http://arxiv.org/abs/2002.12419}{{\tt arXiv:2002.12419 [hep-ph]}}].

\bibitem{ben92}
M.~Beneke, {\it Large-order perturbation theory for a physical quantity},
{\em Nucl.\ Phys.\ B} {\bf 405} (1993) 424.

\bibitem{bro92}
D.J.~Broadhurst,
{\it Large-$N$ expansion of QED: Asymptotic photon propagator and contributions
to the muon anomaly, for any number of loops},
{\em Z.\ Phys.\ C} {\bf 58} (1993) 339.

\bibitem{bkm00}
D.J.~Broadhurst, A.L.~Kataev, and C.J.~Maxwell,
{\it Renormalons and multiloop estimates in scalar correlators, Higgs decay
 and quark-mass sum rule},
{\em Nucl.\ Phys.\ B} {\bf 592} (2001) 247,
[\href{http://arxiv.org/abs/hep-ph/0007152}{{\tt hep-ph/0007152}}].

\bibitem{sc88}
V.P.~Spiridonov and K.G.~Chetyrkin,
{\it Nonleading mass corrections and renormalization of the operators
 $m\bar\psi\psi$ and $G_{\mu\nu}^2$},
{\em  Sov.\ J.\ Nucl.\ Phys.} {\bf 47} (1988) 522,
[{\em Yad.\ Fiz.} {\bf 47} (1988) 818].

\bibitem{st90}
L.R.~Surguladze and F.V.~Tkachov,
{\it Two-loop effects in QCD sum rules for light mesons},
{\em Nucl. Phys. B} {\bf 331} (1990) 35.

\bibitem{bmo18}
D.~Boito, P.~Masjuan and F.~Oliani,
{\it Higher-order QCD corrections to hadronic $\tau$ decays from Pad\'e
 approximants},
JHEP {\bf 08} (2018) 075,
[\href{http://arxiv.org/abs/1807.01567}{{\tt arXiv:1807.01567 [hep-ph]}}].

\bibitem{cap19}
I.~Caprini,
{\it Higher-order perturbative coefficients in QCD from series acceleration
 by conformal mappings},
{\em Phys.\ Rev.\ D} {\bf 100} (2019) 056019,
[\href{http://arxiv.org/abs/1908.06632}{{\tt arXiv:1908.06632 [hep-ph]}}].

\bibitem{jm36}
M.~Jamin, and R.~Miravitllas,
{\it work in progress}.

\bibitem{tvz80}
O.V.~Tarasov, A.A.~Vladimirov, and Yu.A.~Zharkov,
{\it The Gell-Mann-Low function of QCD in the three-loop approximation},
{\em Phys.\ Lett.\ B}, {\bf 93} (1980) 429.

\bibitem{lrv97}
T.~van~Ritbergen, J.A.M.~Vermaseren, and S.A.~Larin,
{\it The four-loop $\beta$-function in quantum chromodynamics},
{\em Phys.\ Lett.\ B}, {\bf 400} (1997) 379,
[\href{http://arxiv.org/abs/hep-ph/9701390}{{\tt hep-ph/9701390}}].

\bibitem{cza04}
M.~Czakon,
{\em The four-loop {QCD} $\beta$-function and anomalous dimensions},
{\em Nucl.\ Phys.\ B}, {\bf 710} (2005) 485,
[\href{http://arxiv.org/abs/hep-ph/0411261}{{\tt hep-ph/0411261}}].

\bibitem{bck16}
P.A.~Baikov, K.G.~Chetyrkin and J.H.~K{\"u}hn,
{\it Five-loop running of the QCD coupling constant},
[\href{http://arxiv.org/abs/arXiv:1606.08659}{\tt arXiv:1606.08659 [hep-ph]}].

\bibitem{vlr97}
J.A.M.~Vermaseren, and S.A.~Larin, and T.~van~Ritbergen,
{\it The four-loop quark mass anomalous dimension and the invariant quark mass},
{\em Phys.\ Lett.\ B}, {\bf 405} (1997) 327,
[\href{http://arxiv.org/abs/hep-ph/9703284}{{\tt hep-ph/9703284}}].

\bibitem{bck14}
P.A.~Baikov, K.G.~Chetyrkin and J.H.~K{\"u}hn,
{\it Quark mass and field anomalous dimensions to ${\cal O}(\alpha_s^5)$},
{\em JHEP} {\bf 1410} (2014) 76,
[\href{http://arxiv.org/abs/arXiv:1402.6611}{\tt arXiv:1402.6611 [hep-ph]}].

\end{thebibliography}
\end{document}